\documentclass[journal]{IEEEtran}
\usepackage{amsbsy}%,color,multicol}
\usepackage{floatflt} % text flows around figures

\usepackage{amsmath}
\usepackage{amssymb}
\usepackage{times}
\usepackage{graphics}
\usepackage{graphicx}
\usepackage{xspace}
\usepackage{paralist} % compact and in-paragraph* lists
\usepackage{setspace} % buggy somehow
\usepackage{xypic}
\xyoption{curve}
\usepackage{latexsym}
\usepackage{theorem}
\usepackage{ifthen}
\usepackage{subfigure}
\usepackage{booktabs}
\usepackage{algorithm}
\usepackage{algorithmic}
\usepackage{color}
\ifCLASSINFOpdf
\else
\fi

\newtheorem{te}{Theorem}[section]

\newtheorem{lemma}{Lemma}[section]

\newcommand{\be}{\begin{equation}}
\newcommand{\ee}{\end{equation}}
\newcommand{\ba}{\begin{array}}
\newcommand{\ea}{\end{array}}
\newcommand{\bee}{\begin{eqnarray*}}
\newcommand{\eee}{\end{eqnarray*}}
\newcommand{\bea}{\begin{eqnarray}}
\newcommand{\eea}{\end{eqnarray}}

\newcommand{\wmin}{w_{min}}
\newcommand{\wmax}{w_{max}}
\newcommand{\II}{\mathbb{I}}
\newcommand{\ZZ}{\mathbb{Z}}
\newcommand{\LL}{\mathbb{L}}
\newcommand{\GG}{\mathbb{G}}
\newcommand{\eps}{\varepsilon}
\newcommand{\skt}{\hat{s}^{k}}
\newcommand{\sk}{s^{k}}
\newcommand{\hk}{\mathbb{H}^{k}}
\newcommand{\hkt}{\widehat{\mathbb{H}}^{k}}
\newcommand{\gk}{g^{k}}
\newcommand{\gkt}{\hat{g}^{k}}
\newcommand{\xik}{\xi_{i}^{k}}
\newcommand{\xjk}{\xi_{j}^{k}}
\newcommand{\fik}{\Phi_{k}}
\newcommand{\tik}{{\widehat{\Phi}}_{k}}
\newcommand{\ty}{{\widehat{y}}_{k}}
\newcommand{\bik}{{\widetilde{\Phi}}_{k}}
\newcommand{\by}{{{\tilde{y}}}_{k}}
\newcommand{\xim}{\xi_{min}^{k}}
\newcommand{\ximax}{\xi_{max}^{k}}

\begin{document}
\title{Distributed second order methods with increasing number of working nodes}

\author{Nata$\check{\mbox{s}}$a Krklec Jerinki\'c,
Du$\check{\mbox{s}}$an Jakoveti\'c,
Nata$\check{\mbox{s}}$a Kreji\'c,
Dragana Bajovi\'c% <-this % stops a space
        \thanks{Part of results of this paper was
        presented at the IEEE Global Conference on Signal
and Information Processing (GlobalSIP), Washington, DC, VA, USA,
Dec. 2016. The work of first three authors is supported by the Serbian Ministry of Education,
Science, and Technological Development, Grant no. 174030. The first three authors are with the
Department of Mathematics and Informatics, Faculty of Sciences, University of Novi Sad,
Trg Dositeja Obradovi\'ca 4, 21000 Novi Sad, Serbia. The fourth author is with the
Department of Power, Electronics and ´
Communication Engineering, Faculty of Technical Sciences, University of Novi Sad,
Trg Dositeja Obradovi\'ca 6, 21000 Novi Sad, Serbia. Authors' emails: natasa.krklec@dmi.uns.ac.rs,
djakovet@uns.ac.rs, natasak@uns.ac.rs, dbajovic@uns.ac.rs.}}

\maketitle

\begin{abstract}
Recently,
an idling mechanism has been introduced
in the context of distributed \emph{first order}
methods for minimization of
a sum of nodes' local convex costs
over a generic, connected network.
With the idling mechanism, each node~$i$,
at each iteration~$k$,
is active -- updates its
 solution estimate and exchanges
  messages with its network neighborhood --
   with probability~$p_k$,
   and it stays idle with probability~$1-p_k$, while
   the activations are independent
   both across nodes and across iterations.
%    The idling mechanism involves an increasing
%    number of nodes in the algorithm (on average)
%    as the iteration counter~$k$ grows,
%    thus avoiding unnecessarily expensive exact updates
%    at the initial iterations while performing
%     beneficial close-to-exact updates near the solution.
      In this paper, we
      demonstrate that the
      idling mechanism can be successfully
     incorporated in \emph{distributed second order methods} also.
      Specifically, we apply
      the idling mechanism to the recently
      proposed Distributed Quasi Newton method~(DQN). We first show theoretically that,
      when $p_k$ grows to one across iterations in a controlled manner,
      DQN with idling exhibits very similar
      theoretical convergence and convergence rates
 properties as the standard DQN method, thus
  achieving the same order of
 convergence rate~(R-linear) as the standard DQN, but  with
 significantly cheaper updates.
 Simulation examples confirm the benefits of
 incorporating the idling mechanism, demonstrate
 the method's flexibility with respect to
 the choice of the~$p_k$'s, and compare the proposed idling method with
 related algorithms from the literature.
\end{abstract}


% \fi
\IEEEpeerreviewmaketitle

\section{Introduction}

\textbf{Context and motivation}.
The problem of distributed minimization
of a sum of nodes' local costs across a (connected) network
has received a significant and growing interest in the past decade, e.g., \cite{nedic_T-AC,SayedEstimation,WotaoYinExtra,ribeiroNNpart1}.
 Such problem arises in various application domains,
including wireless sensor networks, e.g.,~\cite{BoydFusion}, smart grid, e.g.,~\cite{SoummyaSmartGrid},
distributed control applications, e.g.,~\cite{BulloBook}, etc.

In the recent paper~\cite{DGDSN}, a class of novel distributed \emph{first order methods}
 has been proposed,
 motivated by the so-termed ``hybrid'' methods in~\cite{schmidt} for (centralized)
 minimization of a sum $\sum_{i=1}^n f_i(x)$
of convex component functions.
 The main idea underlying a hybrid method in~\cite{schmidt} is that it is
 designed as a combination of
1) an incremental or stochastic gradient method (or, more generally, an incremental or stochastic
Newton-like method) and 2) a full (standard) gradient method
(or a standard Newton-like method); it
behaves as the stochastic method at the initial algorithm stage,
and as the standard method at a later stage.
 An advantage of the hybrid method is that it potentially
inherits some favorable properties of both incremental/stochastic and standard methods, while  eliminating their important drawbacks.
 For example, the hybrid exhibits fast convergence
at initial iterations~$k$ while having inexpensive updates,
just like incremental/stochastic methods. On the other hand, it eliminates the oscillatory behavior of
incremental methods around the solution for large~$k$'s (because
for large~$k$'s it behaves as a full/standard method).
 Hybrid methods calculate
a search direction at iteration~$k$ based on
a subset (sample) of the $f_i$'s,
where the sample size is small at the initial iterations
(mimicking a stochastic/incremental method),
while it approaches the full sample~$n$ for large $k$'s (essentially matching a full, standard method).

%In this context, each component function~$f_i$ belongs to
%a different node~$i$ in a generic, connected network.
% This problem, namely of distributed minimization
%of a sum of nodes' local costs,
%has received a significant attention, e.g., \cite{nedic_T-AC,SayedEstimation,WotaoYinExtra,ribeiroNNpart1},
%and it finds applications in many domains,
%including wireless sensor networks, e.g.,~\cite{BoydFusion}, smart grid, e.g.,~\cite{SoummyaSmartGrid},
%distributed control applications, e.g.,~\cite{BulloBook}, etc.

With distributed first order methods in~\cite{DGDSN},
the sample size at iteration~$k$
 translates into the number of
 nodes that participate in the distributed algorithm at~$k$.
 More precisely, therein we introduce an idling mechanism
 where each node in the network
 at iteration~$k$ is active with probability~$p_k$
  and stays idle with probability~$1-p_k$,
  where $p_k$ is nominally increasing to one with~$k$,
  while the activations are independent
  both across nodes and through iterations.
  Reference~\cite{DGDSN} analyzes convergence rates for
  a distributed gradient method with the idling mechanism
  and demonstrates by simulation that idling brings
  significant communication and computational
  savings.

\textbf{Contributions}. The purpose of this paper is to demonstrate that the idling mechanism
  can be incorporated in distributed \emph{second order, i.e., Newton-like}
   methods also, by both 1) establishing the corresponding convergence rate analytical results,
   and 2) showing through simulation examples that idling
   continues to bring significant efficiency improvements. Specifically,
   we incorporate here the idling mechanism in
   the Distributed Quasi Newton method~(DQN)~\cite{DQN}. (The DQN method has been proposed and analyzed in~\cite{DQN}, only for the scenario when all nodes are active at all times.)
   DQN and its extension
   PMM-DQN are representative distributed second order methods
   that exhibit competitive performance with respect to the current distributed
   second order alternatives, e.g.,~\cite{ribeiroNNpart1,NewtonRaphsonConsensus,DQM}.

Our main results are as follows.
We first carry out a theoretical analysis of the idling-DQN assuming that the
$f_i$'s are twice continuously differentiable with
bounded Hessians. We show that,
as long as $p_k$ converges to
one at least as fast as $1-1/k^{1+\zeta}$
($\zeta>0$ arbitrarily small),
the DQN method with idling
converges in the mean square sense and almost surely to the same point as the standard DQN method
 that activates all nodes at all times. Furthermore,
 when $p_k$ converges to one at a geometric rate,
 then the DQN algorithm with idling
 converges to its limit at a R-linear
 rate in the mean square sense.
  Simulation examples demonstrate that idling can bring to DQN
 significant improvements in computational
 and communication efficiencies.

 We further demonstrate by simulation significant
 flexibility of the proposed idling mechanism
 in terms of tuning of the activation sequence~$p_k$.
 The simulations show that the idling-DQN method
 is effective for scenarios when
 $p_k$ increases but does not eventually converge to one (stays bounded
 away from one), even when $p_k$ is kept constant across iterations.
 The latter two cases are relevant in practice
 when, due to node/link failures or asynchrony,
 the networked nodes do not have a full control over designing the $p_k$'s,
 or when
 an increasing sequence of the $p_k$'s
 may be difficult to implement.
 We also compare by simulation the proposed idling-DQN method
 with a very recent
 distributed second order method with
 randomized nodes' activations in~\cite{ErminWeiAsyncNN}.
  With constant activation probabilities,
  idling-DQN performs very similarly to \cite{ErminWeiAsyncNN},
  while when the activation probabilities are
  tuned as proposed here, idling-DQN performs
  favorably over~\cite{ErminWeiAsyncNN}.

From the technical side, extending the
analysis of either distributed gradient methods
with idling~\cite{DGDSN} or DQN without idling~\cite{DQN}
to the scenario considered here is highly
nontrivial. With respect to
the standard DQN (without idling),
here we need to cope with \emph{inexact
variants} of the DQN-type second order search directions.
 With respect to gradient methods with idling,
 showing boundedness
 of the sequence of iterates and
 consequently bounding the
 ``inexactness amounts'' of the search directions
 are considerably more challenging and require a different approach.

\textbf{Brief literature review}. There has been a significant progress in
the development of distributed second order methods
 in past few years. Reference~\cite{ribeiroNNpart1}
 proposes a method based on a penalty-like interpretation~\cite{cdc-submitted}
  of the problem of interest, by applying
  Taylor expansions of the
  Hessian of the involved penalty function.
  In~\cite{NewtonRaphsonConsensus},
  the authors develop a distributed version
  of the Newton-Raphson algorithm
  based on consensus and time-scale
  separation principles.
  References~\cite{DQM} and~\cite{ESOM}
  propose distributed
  second order methods based
  on the alternating direction method of multipliers
   and the proximal method of multipliers,
   respectively.
   References~\cite{AccelDualAscent,EWnew}
    develop distributed second order methods
    for the problem formulations that are related but different
    than what we consider in this paper, namely they study
    network utility maximization type problems.
   All the above
   works~\cite{ribeiroNNpart1,NewtonRaphsonConsensus,DQN,DQM,ESOM,AccelDualAscent,EWnew}
    assume that
    all nodes are active across all iterations,
    i.e., they are not concerned with
    designing nor analyzing
    methods with randomized nodes' activations.

More related with our work are papers that study distributed first and second order methods
with randomized nodes' or links' activations.
 References~\cite{nedic_T-AC,ASU_Math_Prog} consider distributed first
 order methods with deterministically or randomly varying
 communication topologies.
Reference~\cite{nedic-gossip}
 proposes a distributed gossip-based first order method
 where only two randomly picked nodes are active at each iteration.
 The authors of~\cite{SayedAsync1,SayedAsync2}
 carry out a comprehensive analysis of (first order)
 diffusion methods, e.g.,~\cite{SayedEstimation},
 under a very general model of asynchrony in
 nodes' local computations and communications.
  Reference~\cite{SayedWotaoYinAsync}
   proposes a proximal distributed first order method
  that provably converges to the exact solution
  under a very general model of asynchronous communication and
  asynchronous computation. Other relevant works on first order or alternating direction methods include, e.g.,~\cite{HongADMM,WrightAsync}.

The authors of~\cite{ErminWeiAsyncNN}
 propose an asynchronous version of
  the (second order) Network Newton method in~\cite{ribeiroNNpart1},
  wherein a randomly selected node
  becomes active at a time and performs a Network Newton-type second order update.
   The paper~\cite{RibeiroQN}, see also \cite{RibeiroQN2,RibeiroQN3}, proposes and analyzes an asynchronous distributed
  quasi-Newton method that is based on an asynchronous
  implementation of the {Broyden}–{Fletcher}–{Goldfarb}–{Shanno} (BFGS) matrix update.

Among the works discussed above, perhaps the closest to our randomized activation model
are the models studied in, e.g.,~\cite{SayedAsync1,SayedAsync2,SayedWotaoYinAsync,RibeiroQN}.
However, these works are still very different from ours. While they are primarily concerned
with establishing convergence guarantees under
various asynchrony effects that are not in control of the networked nodes,
 our aim here is to demonstrate that
 a carefully designed, node-controlled ``sparsification'' of
 the workload across network -- as inspired by work~\cite{schmidt} from centralized optimization -- can yield significant savings
 in communication and computation.
  Importantly, and as demonstrated in simulations here,
  significant savings can be achieved even when
  nodes have only a partial control over
  the network-wide workload orchestration,
  due to, e.g., asynchrony, link failures, etc.

Finally, in a companion paper~\cite{GlobalSip},
we presented a brief preliminary
version of the current paper,
wherein a subset of the results
here are presented without proofs.
Specifically,~\cite{GlobalSip}
considers convergence of DQN with idling only when the activation probabilities
geometrically converge to one, while here we also consider
the scenarios where the activation probability converges to
one sub-linearly; we also include here extensions
where this parameter stays bounded away from one or is kept constant (and less than one) across iterations.

\textbf{Paper organization}.
 Section 2 describes the model that we assume
 and gives the necessary preliminaries.
 Section~3 presents the DQN algorithm with idling,
 while Section~4 analyzes its convergence and convergence rate.
 Section~5 considers DQN with idling in the
 presence of persisting idling, i.e.,
 it considers the extension when
 $p_k$ does not necessarily converge to one.
 Section~6 provides numerical examples.
 Finally, we conclude in Section~7.
 %Some lengthy
 %proofs are relegated to the Appendix.

\section{Model and preliminaries}

Subsection~{2.1} gives preliminaries
and explains the network and optimization models
that we assume. Subsection~{2.2}
 briefly reviews the DQN algorithm
 proposed in~\cite{DQN}.

\subsection{Optimization--network model}

We consider distributed optimization where
$n$ nodes in a connected network solve the following unconstrained problem:
\be \label{objective}
\mathrm{min}_{x \in {\mathbb R}^p}\,\,f(x) = \sum_{i=1}^{n} f_i(x).
\ee
Here, $f_i:\,{\mathbb R}^p \rightarrow \mathbb R$ is a convex
function known only by node~$i$.
We impose the following assumptions on the $f_i$'s.

\noindent {\bf Assumption A1.} Each function $f_{i} : \mathbb{R}^p \rightarrow \mathbb{R}, \; i=1,\ldots,n$ is twice continuously differentiable, and there exist constants $0<\mu\leq L<\infty$ such that for every $x \in \mathbb{R}^p$
$$\mu I \preceq \nabla^{2} f_{i}(x) \preceq L I.$$

Here, $I$ denotes the $p \times p$ identity matrix,
and $P \preceq Q$ (where $P$ and $Q$ are symmetric matrices)
means that $Q-P$ is positive semidefinite.
Assumption A1 implies that
each $f_i$ is strongly convex
with strong convexity parameter~$\mu$,
and it also has Lipschitz continuous
gradient with Lipschitz constant~$L$, i.e.,
for all $i=1,...,n$, for every $x,y \in {\mathbb R}^p$, there holds:
\begin{eqnarray*}
&\,&f_i(y) \geq
f_i(x) + \nabla f_i(x)^T (y-x) + \frac{\mu}{2}\|x-y\|^2\\
&\,&\|\nabla f_i(x) - \nabla f_i(y)\|
\leq
L\,\|x-y\|.
\end{eqnarray*}
Here, notation~$\|\cdot\|$
stands for the Euclidean
norm of its vector argument and
the spectral norm of its matrix argument.
Under Assumption A1,
problem~\eqref{objective}
is solvable and has the unique
solution~$\overline{x}^{*} \in {\mathbb R}^p$.

Nodes $i=1,...,n$ constitute an undirected network
$ {\cal G} = ({\cal V},{E}) $,
where $\cal V$ is the set of nodes and
$E$ is the set of edges.
Denote by $m$ the total number
of (undirected) edges (the cardinality
of $E$).
 The presence of edge $\{i,j\} \in E$
  means that the nodes~$i$ and $j$
   can directly exchange messages
   through a communication link.
 Further, let $O_{i}$  be the set of all neighbors of a node $i$ (excluding~$i$), and define also  $\bar{O}_{i}=O_{i}\bigcup \{i\}$.

\noindent {\bf Assumption A2.}
 The network  $ {\cal G} = ({\cal V},{E}) $ is connected, undirected and simple (no self-loops nor multiple links).

We associate with network $\cal G$
 a $n \times n$ weight matrix that has the following properties.
 %(Here $(\cdot)^T$ denotes the matrix transposition.)

 \noindent{\bf Assumption A3.} Matrix $ W = W^T \in \mathbb{R}^{n \times n} $ is stochastic,
 with elements $ w_{ij} $ such that
  \begin{eqnarray*}
  &\,& w_{ij} > 0 \mbox{ if } \{i,j\} \in {E}, \;  w_{ij} = 0 \mbox{ if }  \{i,j\} \notin {E},\,i \neq j,\\
   &\,& \mbox{ and }  w_{ii} = 1 - \sum_{j \in O_i} w_{ij};
  \end{eqnarray*}
 further,  there exist constants $\wmin$ and $\wmax$ such that for $i=1,\ldots,n$, there holds:
 $$0 < \wmin \leq w_{ii} \leq  \wmax <1.$$

 Denote  by $ \lambda_1 \geq \ldots \geq \lambda_n $ the eigenvalues of $ W. $ Then, we have that $ \lambda_1 = 1,$ all the remaining eigenvalues
 of $W$ are strictly less than one in modulus, and
 the eigenvector that corresponds to the unit eigenvalue is
 $ e := \frac{1}{\sqrt{n}}(1,\ldots,1)^T.  $

%For future reference, we introduce the following notation.
Let $ x=(x_1^T,\ldots,x_n^T)^T \in \mathbb{R}^{np} $,
 with $ x_i \in {\mathbb{R}}^p, $ and denote by  $ \ZZ \in \mathbb{R}^{np \times np} $ the
 Kronecker product of $ W $ and the identity $ I \in \mathbf{R}^{p\times p}, \ZZ = W \otimes I. $\footnote{Throughout, we shall use ``blackboard bold'' upper-case letters
 for matrices of size $(n p) \times (np)$ (e.g., $\mathbb{Z}$),
 and standard upper-case letters for matrices of size $n \times n$ or $p \times p$ (e.g., $W$).}
 We will make use of the following  penalty reformulation of~(\ref{objective})~\cite{cdc-submitted}:
  \be \label{reformulation1}
 \mathrm{min}_{x \in {\mathbb{R}^{np}}} \Phi(x) := \alpha \sum_{i=1}^{n} f_i(x_i) + \frac{1}{2} x^T({\mathbb {I}} - \ZZ)x.
 \ee
 Here, $\alpha>0$ is a constant that,
 as we will see ahead, plays the role
 of a {step size} with the distributed algorithms that we consider.
 The rationale behind
 introducing problem~\eqref{reformulation1}
 and function~$\Phi$
 is that they enable one to interpret the
 distributed first order method in~\cite{nedic_T-AC}
  to solve~\eqref{objective}
  as the (ordinary) gradient method
  applied on~$\Phi$,
  which in turn facilitates the development of second order
  methods; see, e.g.,~\cite{ribeiroNNpart1}, for details.
   Denote by $x^* = \left((x_1^*)^T,...,(x_n^*)^T \right)^T \in {\mathbb R}^{np}$
    the solution to~\eqref{reformulation1},
    where $x_i^* \in {\mathbb R}^p$, for all $i=1,...,n$.
    It can be shown that, for all $i=1,...,n$,
    $\|x_i^* - \overline{x}^*\| = O(\alpha)$, i.e.,
     the distance
     from the desired solution~$\overline{x}^*$
      of~\eqref{objective} and $x_i^*$ is
      of the order of {step size}~$\alpha$, e.g.,~\cite{ribeiroNNpart1}.
% The reason behind introducing~\eqref{reformulation1}
% is that it
 Define also $F:\,{\mathbb R}^{np} \rightarrow \mathbb R$,
 $F(x)=\sum_{i=1}^{n} f_{i}(x_{i})$.

%Throughout the paper, we will consider
%iterative distributed algorithms,
%where nodes iteratively update their
%estimates of the solution to~(1)
%by exchanging messages
%with their immediate neighbors only in
%graph~$\mathcal G$. We will consider
%synchronous methods,
%in the sense that
%all nodes perform solution updates
%according to a global iteration counter
% $k=0,1,2,...$

\textbf{Remark}. Under the assumed setting,
function $\Phi$ has Lipschitz
continuous gradient with a Lipschitz constant
that can be taken as
 $L_{\Phi} = \alpha L +2(1-\wmin)$. Moreover,
function~$\Phi(x)$ is strongly convex with
 a strong convexity modulus that can be taken as $\mu_{\Phi} = \alpha\,\mu>0$.

We study distributed
second order algorithms
to minimize function~$\Phi$ (and hence,
find a near optimal solution of~\eqref{objective}).
Therein, the Hessian of function~$\Phi$ and its splitting
 into a diagonal and an off-diagonal part will play an important role.
  Specifically, first consider the splitting:
 \begin{eqnarray}
 \label{eqn-splitting}
W_{d} &=& diag(W),\:\:\:W_u=W-W_{d}\\
\ZZ &=& \ZZ(W)=\ZZ_{d}+\ZZ_{u},\:\:\mathrm{where} \nonumber\\
\ZZ_{d} &=& W_{d} \otimes I=diag(\ZZ), \:\:\mathrm{and}\:\:\ZZ_{u}=W_{u} \otimes I. \nonumber
 \end{eqnarray}
 (Here, $diag(P)$ is the diagonal matrix
 with the diagonal elements equal to those of matrix~$P$.)

%Throughout the paper, we will
%consider both DQN and DQN with idling.
%To ease notation, we will denote the iterates
%of both methods by $x^k =(\,(x_1^k)^T,...,(x_N^k)^T\,)^T \in {\mathbb R}^{n p}$,
%where $x_i^k \in {\mathbb R}^p$ plays the role
%of the solution estimate of node~$i$, $i=1,...,n$, and
%$k=0,1,...$ is the iteration counter;
%it will be clear from context which
%algorithm is in question.
Further, decompose the Hessian of $\Phi$ as:

\be \label{hes} \nabla^2 \Phi(x)= {\mathbb A}(x)  - \GG, \ee
with ${\mathbb A}:\,{\mathbb R}^{np} \rightarrow {\mathbb R}^{(np) \times (np)}$
given by:
\be \label{Ak}
{\mathbb A}(x) =\alpha \nabla^{2} F(x) + (1+\theta)(\II-\ZZ_{d}),
\ee
and
\begin{equation}
\label{eqn-splitting-GG}
\GG=\GG(\ZZ,\theta)=\ZZ_{u}+\theta (\II-\ZZ_{d}),
\end{equation}
for some $\theta \geq 0$.

We close this subsection with the following result that will be needed in subsequent analysis.
% For claim~(a) in Lemma~2.1, see, e.g. Corollary~4
%  in~\cite{Carlson}; claims (b) and (c) follow, e.g.,
%  from Corollary~4 in~\cite{EigenvaluesNew}.
%
%
%\begin{lemma}
%Consider two square Hermitean complex matrices $H_1$ and $H_2$ of equal dimensions.
%\begin{itemize}
%\item [(a)] If $H_2$ is positive semidefinite,
%then the product $H_1 H_2$ has all real eigenvalues.
%\item [(b)] If $H_1$ and $H_2$ are both positive semidefinite,
%then all eigenvalues of $H_1 H_2$ are nonnegative.
%\item [(c)] If $H_1$ and $H_2$ are both positive
%definite, then all eigenvalues
%of $H_1 H_2$ are strictly positive.
%\end{itemize}
%\end{lemma}
For claim~(a), see Lemma~{3.1} in \cite{RamNedicVeeravalli};
 for claim~(b), see, e.g., Lemma~{4.2} in \cite{kkj}.
\begin{lemma} Consider
 a deterministic sequence
  $\{a_k\}$ converging to zero, with $a_k>0$, $k=0,1,...$,
  and let $\nu \in (0,1)$.
  \begin{itemize}
  \item [(a)] Then, there holds:
  \begin{equation}
  \label{eqn-sum-new-100}
  \sum_{t=1}^k
 \nu^{k-t}a_{t-1} \rightarrow 0\mathrm{\,\,as\,\,}k \rightarrow \infty.
  \end{equation}
  \item [(b)]
  If, moreover, $\{a_k\}$
   converges to zero R-linearly,
   then the sum in \eqref{eqn-sum-new-100}
   also converges to zero R-linearly.
   \end{itemize}
\end{lemma}
%The matrix and vector norms that will be used in the sequel are defined here.
%%Let $ \|a\| $ denote the Euclidean norm of vector~$a$.
% Let $\|P\|_2$ denote the spectral norm of matrix~$P$.
%%
%%an arbitrary vector norm on $ \mathbb{R}^{n} $ and the corresponding  matrix norm on $ \mathbb{R}^{n \times n}. $
% Further, for a matrix ${\mathbb M} \in \mathbb{R}^{np\times np}$ with blocks $M_{ij} \in  \mathbb{R}^{p\times p}$, we define
% the following block norm:
%$$\|{\mathbb M}\|:= \max_{j =1,\ldots,n} \sum_{i=1}^{n} \|M_{ij}\|_{2},$$
%where, as noted, $\|M_{ij}\|_{2}$ is the spectral norm of $M_{ij}$.
%With vectors~$a$, we will work only with
%the Euclidean norm, which we denote by~$\|a\|$.
%
%For a vector $x \in \mathbb{R}^{np}$ with blocks $x_{i} \in  \mathbb{R}^{p}$, we use Euclidean norm, i.e.
%$$\|x\|^2:=\sum_{i=1}^{n} \|x_{i}\|^{2}.$$

\subsection{Algorithm DQN}
We will incorporate
the idling mechanism
in the algorithm DQN proposed in~\cite{DQN}.
The main idea behind
DQN is to approximate the
Newton direction
with respect to function~$\Phi$ in \eqref{reformulation1}
 in such a way that
 distributed implementation is
 possible, while the error
 in approximating the Newton direction is not large.
%This is a representative
% distributed second order method
  For completeness,
 we now briefly review DQN.
  All nodes are assumed to be synchronized
according to a global clock and perform in parallel iterations $k = 0, 1, ...$
  The
 algorithm maintains iterates $x^k =(\,(x_1^k)^T,...,(x_n^k)^T\,)^T \in {\mathbb R}^{n p}$,
  over iterations $k=0,1,...$,
where $x_i^k \in {\mathbb R}^p$ plays the role
of the solution estimate of node~$i$, $i=1,...,n$.
 DQN is presented in Algorithm~1 below.
  Therein, ${\mathbb A}_k:={\mathbb A}(x^k)$, where
 ${\mathbb A}(x)$ is given in~\eqref{Ak};
 also, notation $\| \cdot \|$ stands for
 the Euclidean norm of its vector argument
 and spectral norm for its matrix argument.

\vspace{4mm}
 \noindent{\bf Algorithm 1: DQN in vector format}\\
 Given $ x^0 \in \mathbb{R}^{np},  \varepsilon, \rho , \theta, \alpha> 0. $
 %\be \label{del} \delta \in \left(0,\frac{1}{\alpha L+0.5}\right), \quad \rho \in \left(0,2 \lef(\frac{1}{\alpha L+0.5}-\delta\right)\right].\ee
 Set $ k = 0. $
 \begin{itemize}
 \item[(1)] Chose a diagonal matrix $\LL_{k} \in \mathbb{R}^{np \times np} $ such that
 \begin{equation}
 \label{eqn-DQN-1}
 \|\LL_{k}\| \leq \rho.
 \end{equation}
 \item[(2)]
Set
\begin{equation}
\label{eqn-DQN-2}
s^k = - (\II - \LL_k \GG) {\mathbb A}_k^{-1} \nabla \Phi(x^k).
\end{equation}
 \item[(3)] Set
\begin{equation}
\label{eqn-DQN-3}
 x^{k+1} = x^k + \varepsilon s^k, \; k = k+1.
\end{equation}
 \end{itemize}
 We now briefly
 comment on the algorithm
 and the involved parameters. DQN takes the step
 in the direction~$s^k$ scaled
 with a positive {step size}
 $\varepsilon>0$;
 direction~$s^k$ in~\eqref{eqn-DQN-2}
 is an approximation
 of the Newton direction
 \[s_{\mathrm{N}}^k=-\left[\nabla^2 \Phi(x^k)\right]^{-1} \nabla \Phi(x^k).\]
 Unlike Newton direction~$s_{\mathrm{N}}^k$,
 direction~$s^k$
 admits an efficient distributed implementation.
  Inequality~\eqref{eqn-DQN-1} corresponds to a safeguarding step
  that is needed to ensure
  that $s^k$ is a descent direction
  with respect to function~$\Phi$;
  the nonnegative parameter $\rho$
   controls the safeguarding (see~\cite{DQN}
   for details on how to set $\rho$
    for a given problem).
   The diagonal
   matrix~$\mathbb{L}_k$
    controls the off-diagonal part
    of the Hessian inverse approximation~\cite{DQN}.
    Various choices for~$\mathbb{L}_k$
    that are easy to implement and do not induce large extra
    computational and communication costs
    have been introduced in~\cite{DQN}.
    Possible and easy-to-implement
    choices are $\mathbb{L}_k=0$
     and $\mathbb{L}_k = - \rho\,\mathbb{I}$.
    %In this paper,
%    we analyze
%    convergence of DQN
%    with idling for $\mathbb{L}_k = - \rho_k\,\mathbb{I}$,
%    where $\rho_k \in {\mathbb R}$
%     is a (possibly iteration-varying) scalar parameter
%      that satisfies $|\rho_k|\leq \rho$, for all $k=0,1,...$
%      (The results can
%      be extended to
%      generic $\mathbb{L}_k$'s
%      under an additional mild
%      assumption on
%      the idling parameters -- see ahead the Remarks
%      associated with Lemma~{4.1}).
      %

      As it is usually
      the case with
      second order methods,
      step size $\varepsilon$
       should be
       in general strictly
       smaller than one to
       ensure global convergence.
       However,
       extensive
       numerical
       simulations on
       quadratic and logistic losses
       demonstrate that
       both DQN and
       DQN with idling
       converge globally
       with the full step size $\varepsilon=1$
        and choices
        $\mathbb{L}_k=0$, and $\mathbb{L}_k=-\mathbb{I}$.
      %
%
%
%    carry out a worst case analysis
%    allowing for generic~$\mathbb{L}_k$'s
%    that obey the safeguarding condition~\eqref{eqn-DQN-1}.

The remaining
  algorithm parameters are as follows.
  Quantity~$\theta \geq 0$
   controls the splitting~\eqref{Ak};
   reference~\cite{DQN}
   shows by simulation that it is usually
   beneficial to adopt a small positive value of~$\theta$
   or $\theta=0$.
   Finally, $\alpha>0$
    defines the penalty
    function~$\Phi$ in \eqref{reformulation1}
     and results in the following tradeoff
     in the performance of DQN:
     a smaller value of $\alpha$
      leads to a better asymptotic
      accuracy of the algorithm, while
      it also slows down the algorithm's
      convergence rate.
      By asymptotic accuracy,
      we assume here the distance between the point of convergence of DQN~$x^{*}$
      (which is actually equal to the
      solution of~\eqref{reformulation1} -- see~\cite{DQN})
      and the solution $\overline{x}^{*}$ of~\eqref{objective}.
       (As noted before, the corresponding distance
      is~$O(\alpha)$~\cite{ribeiroNNpart1}.)

In Algorithm~2,
we present DQN from the
perspective of distributed implementation.
Therein, we denote by
 $\Lambda_i^k$ and $A_i^k$, respectively,
 the $p \times p$ block on the $(i,i)$-th
 position of matrices
 $\mathbb{L}_k$ and $\mathbb{A}_k$.
 Similarly,
 $G_{ij}$ is the $p \times p$
  block at the $(i,j)$-th position of~$\GG$.
%For the sake of clarity, the proposed algorithm,
%from the perspective of each node~$i$ in the network, is presented in Algorithm~2.
%Denote by
%$L_i^k$ the $i$-th $p \times p$ block of $\LL_k$ (which
%corresponds to node~$i$).

%\begin{algorithm}
%\label{algorithm-general-DQN}

\vspace{4mm}
\noindent{\bf Algorithm 2: DQN -- distributed implementation}\\
\noindent{At each node~$i$, require $ x_i^0 \in \mathbb{R}^{p},  \varepsilon, \rho , \theta, \alpha> 0. $
\begin{itemize}
\item[(1)] Initialization: Each node $i$ sets $k=0$ and $x_i^0 \in {\mathbb R}^p$.
\item[(2)] Each node $i$ transmits $x_i^k$ to all its neighbors
    $j \in O_i$ and receives $x_j^k$ from all $j \in O_i$.
    \item[(3)] Each node $i$ calculates
    \[
    d_i^k = \left( A_i^k\right)^{-1} \left[ \, \alpha\, \nabla f_i(x_i^k)
    + \sum_{j \in O_i} w_{ij} \left( x_i^k - x_j^k\right)\,\right].
    \]
    \item[(4)] Each node~$i$ transmits $d_i^k$ to all its neighbors
    $j \in O_i$ and receives $d_j^k$ from all $j \in O_i$.
    \item[(5)] Each node $i$ chooses a diagonal $p \times p$ matrix $\Lambda_i^k$, such that
    $ \|\Lambda_i^k\| \leq \rho. $
    \item[(6)] Each node $i$ calculates:
    \[
    s_i^k  = - d_i^k + \Lambda_i^k \sum_{j \in \bar{O}_i} G_{ij} \,d_j^k.
    \]
    \item[(7)] Each node $i$ updates its solution estimate as:
    \[
    x_i^{k+1} = x_i^k + \varepsilon\,s_i^k.
    \]
    \item[(8)] Set $k=k+1$ and go to step 2.
\end{itemize}

Note that, when $\Lambda_i^k \equiv 0$, for all $i,k$,
steps 4-6 are skipped,
and the algorithm involves
a single communication round per iteration, i.e.,
a single transmission of a $p$-dimensional
vector (step~2) by each node;
when $\Lambda_i^k \equiv - \rho_k\, I$,
 $\rho_k \neq 0$, then two communication rounds
 (steps 2 and 4) per~$k$ are involved.

\section{Algorithm DQN with idling}

Subsection~{3.1}
explains the idling mechanism,
while Subsection~{3.2}
 incorporates this mechanism in the DQN method.

\subsection{Idling mechanism}

 We incorporate in DQN the following idling mechanism.
Each node $i$, at each iteration~$k$,
is active with probability~$p_k$, and it
is inactive with probability~$1-p_k$.\footnote{We continue to assume that all nodes are synchronized
according to a global iteration counter~$k = 0, 1, ...$}
Active nodes perform updates of their
solution estimates $(x_i^k)$'s
and participate in each communication round
of an iteration, while
inactive nodes do not perform
any computations nor communications, i.e.,
their solution estimates
$(x_i^k)$'s
 remain unchanged.
  Denote by $\xi_i^k$
   the Bernoulli
   random variable that governs
   the activity of node $i$
    at iteration $k$.
    Then, we
    have that probability
    ${P}\left( \xi_i^k=1 \right) =
    1-{P}\left( \xi_i^k=0 \right) =p_k$, for all~$i$.
    We furthermore assume that
     $\xi_i^k$ and $\xi_j^{\ell}$
      are mutually independent
      over all $i \neq j$ and
      $k \neq \ell$.
      Throughout the paper, we
      impose the following Assumption on
      sequence~$\{p_k\}$.

      \noindent {\bf Assumption A4.}
      Consider the sequence of activation probabilities~$\{p_k\}$.
      We assume that $p_k \geq p_{min}$, for all $k$,
      for some $p_{min}>0.$
      %That is, $\{p_k\}$ is uniformly bounded from
%      below by $p_{min}$.
       Further, $\{p_k\}$ is a
       non-decreasing sequence
       with $\lim_{k \rightarrow \infty} p_k=1$.
       Moreover,
       we assume that
       \begin{equation}
\label{eqn-u-k-cond}
0 \leq u_k \leq \frac{{\mathcal{C}_{\mathrm{u}}}}{(k+1)^{1+\zeta}},
\end{equation}
where $u_k=1-p_k$, ${\mathcal{C}_{\mathrm{u}}}$ is a positive constant and
$\zeta>0$ is arbitrarily small.

       Assumption A4 means that, on average,
       an increasing number of nodes
       becomes involved in the optimization
       process; that is, intuitively,
       in a sense
       the precision of the optimization process
       increases with the increase of the iteration counter~$k$.
        (Extensions to the scenarios
        when $p_k$ does not necessarily
        converge to one is provided in Section~5.)
        We also assume
        that $p_k$ converges to one
        sufficiently fast, where sublinear convergence
        $1-1/k^{1+\zeta}$ is sufficient.
        %
       %
      %  More precisely, we
      % make the following assumption on the $p_k$'s.

%\noindent {\bf Assumption A4.}
%The sequence $p_k:=1-u_k$, $k=0,1,...,$ is non-decreasing,
%and it is uniformly lower bounded by
%constant~$p_{min}>0$, i.e.,
%$p_k \geq p_{min}$, for all~$k$.
%Moreover, for the sequence $\{u_k\}$, it holds
%that there exists constant ${\mathcal{C}_{\mathrm{u}}} \in (0,+\infty)$,
%such that, for all $k=0,1,...$:
%%
%\begin{equation}
%\label{eqn-u-k-cond}
%0 \leq u_k \leq \frac{{\mathcal{C}_{\mathrm{u}}}}{(k+1)^{1+\zeta}},
%\end{equation}
%%
%where $\zeta>0$ is arbitrarily small.
%
%Assumption~{A4} says that $p_k$
%should converge to one
%(at least) sublinearly,
%with the rate at least $1/k^{1+\zeta}$.
% Note that, under Assumption~{A4},
% sequence~$\{ u_k\}$ is summable.

%For each node, the probability of being activated at iteration $k$ is $p_{k}$. We assume that $p_{k}$ is bounded away from zero, i.e. $p_{k}\geq p_{min}>0$ for every $k=0,1.\ldots$. Random variable that determines the state (active/inactive) of node $i$ at iteration $k$ is $\xi_{i}^{k}$ that has Binomial distribution with parameters $(1,p_{k})$. Node $i$ is active at iteration $k$ if $\xi_{i}^{k}=1$.

For future reference, we also
define the diagonal $(n p) \times (n p)$ (random) matrix ${{{\mathbb{Y}_k}}}= diag(\xi_{1}^{k},\ldots, \xi_{n}^{k}) \otimes I$, where $I$ is the $p \times p$ identity matrix.
Also,
we define the $p \times p$
random matrix $W^k=[w^{k}_{ij}]$ by $w^{k}_{ij}=w_{ij} \xi_{i}^{k} \xi_{j}^{k}$ for $i \neq j$, and $w^{k}_{ii}=1-\sum_{i \neq j} w^{k}_{ij} $.
 Further,
 we let $\ZZ^k:= W^k \otimes I$, and,
 analogously to
 \eqref{eqn-splitting} and \eqref{eqn-splitting-GG},
 we let:
\begin{eqnarray}
\label{eqn-splitting-idling}
W_{d}^k &=& diag(W^k),\:\:\:W_u^k=W^k-W_{d}^k\\
\ZZ^k &=& \ZZ(W^k)=\ZZ_{d}^k+\ZZ_{u}^k,\:\:\mathrm{where} \nonumber\\
\ZZ_{d}^k &=& W_{d}^k \otimes I=diag(\ZZ^k), \:\:\mathrm{and}\:\:\ZZ_{u}^k=W_{u}^k \otimes I. \nonumber\\
\GG^k &=& \GG(\ZZ^k,\theta)=\ZZ_{u}^k+\theta (\II-\ZZ_{d}^k).
 \end{eqnarray}
%
%$\ZZ^{k}=\ZZ(W^{k})$ and we define $\ZZ^{k}_{d}$ and $\ZZ^{k}_{u}$ analogously to deterministic case stated above. The same is true for $\GG^{k}$.

Notice that $w^{k}_{ii}=1-\sum_{i \neq j} w_{ij} \xi_{i}^{k} \xi_{j}^{k}\geq 1-\sum_{i \neq j} w_{ij}=w_{ii}\geq w_{min}. $ Further, recall ${\mathbb A}: \,{\mathbb R}^{np} \rightarrow {\mathbb R}^{(np) \times (np)}$
 in~\eqref{Ak}. Using results from~\cite{DQN}, we obtain the following important bounds:\footnote{Throughout subsequent analysis,
we shall state several
relations (equalities and inequalities)
that involve random variables.
These relations hold either
surely (for every outcome), or in expectation. E.g.,
relation \eqref{s1} holds surely.
It is clear from notation which of the two cases is in force.
 Also, auxiliary constants that arise
 from the analysis will be frequently denoted by the capital calligraphic
 letter $\mathcal{C}$ with a subscript that
 indicates a quantity related with the constant in question;
 e.g., see ${\mathcal{C}_{\mathrm{A}}}$ in \eqref{s2}.} %
 \begin{eqnarray}
\label{s2}
\|{\mathbb A}^{-1}(x) \| &\leq& (\alpha \mu +(1+\theta)(1-\wmax))^{-1}\\
&:=&{\mathcal{C}_{\mathrm{A}}},\,\,x  \in {\mathbb R}^{n p} \nonumber \\
\label{s1} \|\GG^{k} \| &\leq& (1+\theta) (1-w_{min})\\
&:=&{\mathcal{C}_{\mathrm{G}}},\,\,k=0,1,... \nonumber \\
 \label{s4} \left\|\GG^{k} {\mathbb A}^{-1}(x) \right\| &\leq& \frac{(1+\theta) (1-w_{min})}{\alpha \mu +(1+\theta)(1-\wmin)},\\
 &\,&\,x  \in {\mathbb R}^{n p},\,\,k=0,1,...\nonumber %:={\mathcal{C}_{\mathrm{G,A}}},\,\,k=0,1,...,\,x\in {\mathbb R}^{np}.
\end{eqnarray}
More generally, for ${\mathbb A}(x)$ there holds:
\begin{eqnarray}
&\,& (\alpha \mu +(1+\theta) (1-\wmax))\II \preceq {\mathbb A}(x) \nonumber  \\
&\,& \preceq (\alpha L +(1+\theta) (1-\wmin))\II,
\,\,x\in {\mathbb R}^{np}. \label{s3}
\end{eqnarray}

Also, notice that $\|{{{\mathbb{Y}_k}}}\| \leq 1$, and $\ZZ^{k} \preceq \II$, for every $k$.

\subsection{DQN with idling}
%In this section we introduce a class of Quasi Newton methods for solving (\ref{reformulation1}). In general, it takes the following form. Let $\LL_{k} \in \mathbb{R}^{np \times np}$ be a diagonal matrix composed of diagonal $ p \times p $ matrices $ \Lambda_i^k, \; i=1,\ldots,n.  $ The method is defined by
%\be \label{defdqnsn} x^{k+1}=x^k+\varepsilon s^k\ee
%where  $ \varepsilon $ is a step size and $s^{k}$ is an approximation of the Newton direction given by
%$$  s^k = - (\II - \LL_k \GG^{k}) {\mathbb A}_k^{-1} ({\frac{\alpha}{p_{k}}} {{{\mathbb{Y}_k}}} \nabla F(x^{k})+(\II-\ZZ^{k})x^{k}). $$

%For the sake of clarity, the proposed algorithm,
%from the perspective of each node~$i$ in the network, is presented in Algorithm~2.
%Denote by
%$L_i^k$ the $i$-th $p \times p$ block of $\LL_k$ (which
%corresponds to node~$i$).

%\begin{algorithm}
%\label{algorithm-general-DQN}

We now incorporate the idling mechanism in the DQN method.
To avoid notational clutter,
we continue to denote by $x^k = \left((x_1^k)^T,...,(x_n^k)^T \right)^T$
 the algorithm iterates, $k=0,1,...$,
 where $x_i^k$
  is node $i$'s estimate
  of the solution to~\eqref{objective} at
  iteration~$k$.
  DQN with idling operates as follows.
  If the
  activation variable $\xi_i^k=1$,
  node $i$ performs an update;
  else, if $\xi_i^k=0$,
  node $i$ stays idle and lets
  $x_i^{k+1}=x_i^k.$
  The algorithm is presented in Algorithm~3 below.
  Therein,
  ${\mathbb A}_k:={\mathbb A}(x^k)$,
  and $A_i^k$ is the $p \times p$ block
  at the $(i,i)$-th position in ${\mathbb A}_k,$
   while $G_{ij}^k$
   is the $p \times p$ block
   of $\GG^k$ at the $(i,j)$-th position.

\vspace{4mm}
\noindent{\bf Algorithm~3: DQN with idling -- distributed implementation}\\
\noindent{At each node~$i$, require $ x_i^0 \in \mathbb{R}^{p},  \varepsilon, \rho , \theta, \alpha> 0, \{p_k\}.$
\begin{itemize}
\item[(1)] Initialization: Node $i$ sets $k=0$ and $x_i^0 \in {\mathbb R}^p$.
\item[(2)] Each node~$i$ generates $\xi_i^k$; if $\xi_i^k=0$, node~$i$ is idle, and goes to step~9;
else, if $\xi_i^k=1$, node~$i$ is active and goes to step~3;
all active nodes do steps 3-8 below in parallel.
\item[(3)] (Active) node $i$ transmits $x_i^k$ to all its active neighbors
    $j \in O_i$ and receives $x_j^k$ from all active $j \in O_i$.
    \item[(4)] Node $i$ calculates
    \[
    d_i^k = \left( A_i^k\right)^{-1} \left[ \, {\frac{\alpha}{p_{k}}} \, \nabla f_i(x_i^k)
    + \sum_{j \in O_i} w^{k}_{ij} \left( x_i^k - x_j^k\right)\,\right].
    \]
    \item[(5)] Node~$i$ transmits $d_i^k$ to all its active neighbors
    $j \in O_i$ and receives $d_j^k$ from all the active $j \in O_i$.
    \item[(6)] Node $i$ chooses a diagonal $p \times p$ matrix $\Lambda_i^k$, such that
    $ \|\Lambda_i^k\| \leq \rho. $
    \item[(7)] Node $i$ calculates:
    \[
    s_i^k  = - d_i^k + \Lambda_i^k \sum_{j \in \bar{O}_i} G^{k}_{ij} \,d_j^k.
    \]
    \item[(8)] Node $i$ updates its solution estimate as:
    \[
    x_i^{k+1} = x_i^k + \varepsilon\,s_i^k.
    \]
    \item[(9)] Set $k=k+1$ and go to step~2.
\end{itemize}

We make a few remarks on Algorithm~3.  First, note that, unlike Algorithm~2, the iterates
$x_i^k$ with Algorithm~3 are random variables. (The initial iterates
$x_i^0$, $i=1,...,n$, in Algorithm~3 are assumed deterministic.)
 Next, note that we implicitly assume that
   all nodes have agreed beforehand
   on scalar parameters
   $\varepsilon, \rho , \theta, \alpha$;
    this can actually be achieved in a distributed way
    with a low communication and computational overhead
    (see Subsection~{4.2} in~\cite{DQN}).
    Nodes also agree beforehand on
   the sequence of
   activation probabilities
   $\{p_k\}$. In other words,
   sequence $\{p_k\}$ is assumed to be available
   at all nodes. For example,
   as discussed in more detail
   in Sections~4 and~6,
   we can let $p_k=1-\sigma^{k+1}$,
   $k=0,1,...$, where $\sigma \in (0,1)$
    is a scalar parameter known by all nodes.
     As each node is aware of the global iteration counter~$k$,
     each node is then able to implement the latter formula for~$p_k$.
      The nodes' beforehand agreement on $\sigma$ can be achieved
      similarly to the agreement on
      other parameters
   $\varepsilon, \rho , \theta, \alpha$~\cite{DQN}.
 Tuning of parameter~$\sigma$  is discussed further ahead.

Parameters $\epsilon$, $\rho$,
and $\alpha$, and
the diagonal matrices $\Lambda_i^k$
 play the same role as in DQN.
 An important difference
 with respect to
 standard DQN appears in step~4,
 where the local \emph{active} node $i$'s
  gradient contribution is
  $\frac{\alpha}{p_k}\nabla f_i(x_i^k)
  = \frac{\alpha}{p_k}\,\xi_i^k\,\nabla f_i(x_i^k)$,
  while with standard DQN
  this contribution equals~$\alpha\,\nabla f_i(x_i^k)$.
   Note that
   the division by $p_k$
    for DQN with idling makes
    the terms of the two
    algorithms balanced \emph{on average},
    because $E[\xi_i^k]=p_k$.

Using notation~\eqref{eqn-splitting-idling},
we represent DQN with idling in Algorithm~4
 in a vector format.
 (Therein,
 $\LL_k = diag\left(
 \Lambda_1^k,...,\Lambda_n^k\right)$.)

%{\bf{Remark}} Inactive node does not update its decision vector, i.e. if a node $i$ is inactive at iteration $k$ then $x_{i}^{k+1}=x_{i}^{k}$.

\vspace{4mm}
 \noindent{\bf Algorithm~4: DQN with idling in vector format}\\
 Given $ x^0 \in \mathbb{R}^{np},  \varepsilon, \rho , \theta, \alpha> 0, \{p_k\}$.
 %\be \label{del} \delta \in \left(0,\frac{1}{\alpha L+0.5}\right), \quad \rho \in \left(0,2 \left(\frac{1}{\alpha L+0.5}-\delta\right)\right].\ee
 Set $ k = 0. $
 \begin{itemize}
 \item[(1)] Chose a diagonal matrix $\LL_{k} \in \mathbb{R}^{np \times np} $ such that $$\|\LL_{k}\| \leq \rho.$$
 \item[(2)]
Set
$$  s^k = - (\II - \LL_k \GG^{k})\, {\mathbb A}_k^{-1} \left({\frac{\alpha}{p_{k}}} {{{\mathbb{Y}_k}}} \nabla F(x^{k})+(\II-\ZZ^{k})\,x^{k}\right). $$
 \item[(3)] Set
 $$ x^{k+1} = x^k + \varepsilon s^k, \; k = k+1. $$
 \end{itemize}

\section{Convergence analysis}
In this Section,
we carry out convergence
and convergence rate analysis
of DQN with idling.
We have two main results,
Theorems~{IV.4} and~{IV.5}.
The former result states
that, under Assumptions A1-A4, DQN with idling
converges to the solution~$x^*$ of~\eqref{reformulation1}
 in the mean square sense and almost surely.
   We then
 show that, when
 activation probability~$p_k$
  converges to one at a geometric rate,
  the mean square convergence
  towards $x^{*}$
   occurs at a R-linear rate.
 Therefore,
 the order of convergence (R-linear rate)
  of the DQN method is preserved
  despite the idling.
  We note that the above result
  does not explicitly
  establish
  computational and communication savings
  with respect to standard DQN.
  An explicit quantification of
  these savings is very challenging
  even for distributed
  first order methods~\cite{DGDSN},
   and even more so here.
However, the theoretical results
in the current section are complemented
in Section~6 with numerical examples;
 they demonstrate that
 communication and computational savings
 usually occur in practice.

The analysis is organized as follows. In Subsection~{4.1},
we relate the search direction of DQN with idling
and the search direction of DQN,
where the former is viewed as an inexact version of the latter.
  Subsection~{4.2} establishes the mean square
  boundedness of the iterates of
  DQN with idling and its implications
    on the ``inexactness'' of search directions.
    Finally, Subsection~{4.3} makes use
    of the results in Subsections~{4.1} and~{4.2}
    to prove the main results on the convergence and
    convergence rate of DQN with idling.

\subsection{Quantifying inexactness of search directions}
We now analyze how ``inexact''
 is the search direction of the DQN with idling
 with respect to the search direction
 of the standard DQN.
  For $x^k$ the iterate of
   DQN with idling, denote by $\skt$
    the search direction
    as with the standard DQN evaluated at $x^k$, i.e.:
  %the search direction proposed therein by $\skt$, i.e.
  %
\be \label{sktilda} \skt=- (\II - \LL_k \GG) {\mathbb A}_k^{-1} (\alpha \nabla F(x^{k})+(\II-\ZZ)x^{k}). \ee
%
%[*** PREDLOG DA SE MOZDA $\skt$ I $\gkt$
% ZAMENE NOTACIJOM $\widehat{s}^{\,k}$ and $\widehat{g}^{\,k}$
% I SLICNO ZA $\tilde{H}^k$.]

Then, the search direction~$\sk$ of DQN with idling
can be viewed as an approximation, i.e., an inexact version, of $\skt$.
  We will show ahead that the error of this approximation is controlled by the activation probability $p_{k}$. In order to simplify notation in the analysis,
  we introduce the following quantities:
   \begin{eqnarray*}
   \hk &=& \II - \LL_k \GG^{k}  \\
   \hkt &=& \II - \LL_k \GG\\
   \gk &=& {\mathbb A}_k^{-1} ({\frac{\alpha}{p_{k}}} {{{\mathbb{Y}_k}}} \nabla F(x^{k})+(\II-\ZZ^{k})x^{k})
   \\
   \gkt &=& {\mathbb A}_k^{-1} (\alpha \nabla F(x^{k})+(\II-\ZZ)x^{k}).
   \end{eqnarray*}
Therefore, $\skt=-\hkt \gkt$ and $\sk=-\hk \gk$.
   Notice that $\|\hk\| \leq 1+\rho {\mathcal{C}_{\mathrm{G}}}:={\mathcal{C}_{\mathrm{H}}}$ and that the same is true for $\hkt$.
   We have the following result on
   the error in
   approximating~$\skt$ with~$\sk$.
   In the following,
   we denote by either $a_i$ or $[a]_i$ the $i$-th $p\times 1$
    block of a $(np)$-dimensional
    vector~$a$; for example,
    we write $\gk_i$ for the $i$-th
    $p$-dimensional block of~$\gk$.

In the following Theorem,
claims~\eqref{eqn-lambda-min-2} and \eqref{eqn-lambda-3}
are from Theorem~{3.2} in~\cite{DQN};
 claim~\eqref{eqn-lambda-min}
  is a straightforward generalization of~\eqref{eqn-lambda-min-2},
  mimicking the proof steps of Theorem~{3.2} in~\cite{DQN};
  hence, proof details are omitted.
%
%  (These results were applied
%  in~\cite{DQN} to the iterates of DQN,
%  but they clearly hold for the iterates
%  of DQN with idling as well;
%  the latter versions of the results are used here.)
  %
\begin{te} \label{tdqn} \cite{DQN}
Let assumptions A1-A3 hold.
Further, let, $\rho \in [0, {\overline{\rho}}_{DQN}]$ where
   \begin{eqnarray*}
   &\,& {\overline{\rho}}_{DQN}= \frac{\alpha \mu +(1+\theta)(1-\wmax)}{(1-\wmin)(1+ \theta)} \\
   &\,& \times \left(\frac{1}{\alpha L+(1+\theta)(1-\wmin)}-\delta\right),
   \end{eqnarray*}
   for some constant~$\delta \in (0,1/(\alpha L+(1+\theta)(1-\wmin)))$.
   Consider random matrices
   ${\mathbb R}_k:= {\mathbb H}^k {\mathbb A}_k^{-1}$
   and
   $\widehat{{\mathbb R}}_k:= \widehat{{\mathbb H}}^k {\mathbb A}_k^{-1} $.
   Then, there holds:
    \begin{eqnarray}
    \label{eqn-lambda-min}
   \lambda_{min} \left(\frac{{\mathbb R}_k+{\mathbb R}_k^T}{2}\right) &\geq&  \delta\\
   \label{eqn-lambda-min-2}
   \lambda_{min} \left(\frac{\widehat{{\mathbb R}}_k + \widehat{{\mathbb R}}_k^T}{2}\right) &\geq&  \delta,
\end{eqnarray}
where $\lambda_{min}(\cdot)$ denotes the minimal eigenvalue.
Moreover, quantity $ \skt $ in~(\ref{sktilda})  satisfies
 the following bounds:
 \begin{equation}
 \label{eqn-lambda-3}
 \nabla^T \Phi(x^k) \skt \leq - \delta \|\nabla \Phi(x^k)\|^2 \; \mbox{and} \; \|\skt \| \leq \beta \|\nabla \Phi(x^k)\|,
 \end{equation}
 where constant~$ \beta=\frac{1+\rho (1+\theta) (1-\wmin)}{\alpha \mu +(1+\theta) (1-\wmax)}$.
 \end{te}
Moreover, it was shown in~\cite{DQN} that $\delta < \beta$, which we use in the proof of the subsequent result.
Furthermore, using the Mean value theorem and Lipschitz continuity of $\nabla \Phi$ we obtain (see the proof of Theorem 3.3 in \cite{DQN} for instance):
%
%{\bf{***Ovde je bila greska, pisalo je sktilda umesto sk}. Ispravila sam.}
%
\begin{eqnarray}
\label{21}
&\,& \Phi\left(x^{k}+\varepsilon\,s^k \right) - \Phi(x^*)\leq \Phi(x^k) \\
&\,&- \Phi(x^*) + \frac{1}{2} \varepsilon^2 {L_{\Phi}} \|s^k\|^2 +  \varepsilon \nabla^T \Phi(x^k) s^k,
\end{eqnarray}
where we recall that $x^{*}$ is the (unique) solution of~(\ref{reformulation1}) and
${L_{\Phi}}$ is a Lipschitz gradient continuity parameter of $\Phi$, which equals ${L_{\Phi}}=\alpha L +2(1-\wmin)$.
Next, we prove that DQN with idling exhibits a kind of nonmonotone behavior where
the ``nonmonotonicity'' term depends on the difference between the search directions $\sk$ and $\skt$.
 %Recall that ${L_{\Phi}}=\alpha L +2(1-\wmin).$

\begin{te} \label{t3}  Let assumptions A1-A3 hold,  $\rho \in [0, {\overline{\rho}}_{DQN}]$ and
$\eps \leq {\overline{\varepsilon}_{DQN}}$, where
\be \label{s22} {\overline{\varepsilon}_{DQN}}= \frac{\delta-q}{2 {L_{\Phi}} (\beta^2+q^2)}, \quad q \in (0,\delta).\ee
  %Moreover, assume that $\LL_{k}= -\rho_k \II $.
 Then
   $$\Phi(x^{k+1}) - \Phi(x^*)\leq (\Phi(x^k) - \Phi(x^*)) \nu(\eps) +e_{k},$$
   where $\nu(\eps) \in (0,1)$ is a constant, and $e_{k}=(\eps^2 {L_{\Phi}}+\eps / q) \|\sk-\skt\|^2.$
 \end{te}
 {\em Proof.} We start by considering $s^k$ in (\ref{21}) and using the bounds from Theorem \ref{tdqn}, i.e.:
 \begin{eqnarray}
&\,&\Phi(x^{k+1}) - \Phi(x^*)  \leq  \Phi(x^k) - \Phi(x^*) \nonumber \\
&+& \frac{1}{2} \varepsilon^2 {L_{\Phi}} \|s^k\pm\skt\|^2 +  \varepsilon \nabla^T \Phi(x^k) (s^k\pm\skt) \nonumber\\
& \leq  &  \Phi(x^k) - \Phi(x^*) + \varepsilon^2 {L_{\Phi}} \|s^k-\skt\|^2+ \varepsilon^2 {L_{\Phi}} \|\skt\|^2+ \nonumber \\
&+&  \varepsilon \nabla^T \Phi(x^k) (s^k-\skt) + \varepsilon \nabla^T \Phi(x^k) \skt \nonumber\\
& \leq & \Phi(x^k) - \Phi(x^*)+ \varepsilon^2 {L_{\Phi}} \|s^k-\skt\|^2 \nonumber \\
&+& \varepsilon^2 {L_{\Phi}} \beta^2 \|\nabla \Phi(x^k)\|^2+ \nonumber \\
&+&  \varepsilon \|\nabla \Phi(x^k)\| \|(s^k-\skt)\| - \varepsilon  \delta \|\nabla \Phi(x^k)\|^2 \label{s23}
 \end{eqnarray}
We distinguish two cases. First, assume that $\|s^k-\skt\| \leq q \|\nabla \Phi(x^k)\|$. In this case, (\ref{s23}) implies that
$$\Phi(x^{k+1}) - \Phi(x^*)  \leq \Phi(x^k) - \Phi(x^*)+ \varphi(\eps)\|\nabla \Phi(x^k)\|^2$$ where
$\varphi(\eps)=\varepsilon^2 {L_{\Phi}}(\beta^2+q^2)-\eps (\delta-q)$. Notice that $\varphi$ is convex, $\varphi(0)=0$ and it attains its minimum at ${\overline{\varepsilon}_{DQN}}$ given in (\ref{s22}) with $\varphi({\overline{\varepsilon}_{DQN}})=-((\delta-q)^2)/(4{L_{\Phi}}(\beta^2+q^2) )$. This also implies that $\varphi(\eps)$ is negative for all $\eps \in (0,{\overline{\varepsilon}_{DQN}}]$. Moreover, $\Phi$ is strongly  convex and there holds $\Phi(x^k) - \Phi(x^*) \leq \frac{1}{{\mu_{\Phi}}}\|\nabla \Phi(x^k)\|^2$ where ${\mu_{\Phi}}=\alpha \mu$ is the strong
convexity parameter. Putting all together we obtain
\be \label{s24} \Phi(x^{k+1}) - \Phi(x^*) \leq (\Phi(x^k) - \Phi(x^*)) \nu(\eps)\ee
where $\nu(\eps)=1+{\mu_{\Phi}} \varphi(\eps)$. Notice that $\nu(\eps)<1$ for all $\eps \in (0,{\overline{\varepsilon}_{DQN}}]$. Moreover,
$$\nu(\eps) \geq 1+{\mu_{\Phi}} \varphi({\overline{\varepsilon}_{DQN}}) =1-\frac{{\mu_{\Phi}} (\delta-q)^2}{4 {L_{\Phi}}(\beta^2+q^2)}.$$
  As ${\mu_{\Phi}} =\alpha \mu < \alpha L < {L_{\Phi}}$ and $(\delta-q)^2 <\delta^2<\beta^2 <\beta^2+q^2$ we conclude that $
\nu(\eps)$ is positive. Therefore,  for all $\eps \in (0,{\overline{\varepsilon}_{DQN}}]$,   (\ref{s24}) holds with $
\nu(\eps)\in (0,1)$.

Now, assume that $\|s^k-\skt\| > q \|\nabla \Phi(x^k)\|$, i.e. $ \|\nabla \Phi(x^k)\| < \|s^k-\skt\| / q$. Together with (\ref{s23}), this implies
\be \label{s25} \Phi(x^{k+1}) - \Phi(x^*)  \leq \Phi(x^k) - \Phi(x^*)+ \hat{\varphi}(\eps)\|\nabla \Phi(x^k)\|^2+e_{k}\ee where
$\hat{\varphi}(\eps)=\varepsilon^2 {L_{\Phi}}\beta^2-\eps \delta$ and  $e_{k}=(\eps^2 {L_{\Phi}}+\eps / q) \|\sk-\skt\|^2.$ The function
$\hat{\varphi}$ has similar characteristics as $\varphi$ and it retains its minimum at ${{\varepsilon}^{\prime}}=\delta/(2 {L_{\Phi}} \beta^2)$ with
$\hat{\varphi}({{\varepsilon}^{\prime}})=-\delta^2/(4 {L_{\Phi}} \beta^2)$. Since ${\overline{\varepsilon}_{DQN}} \leq {{\varepsilon}^{\prime}}$, $\hat{\varphi}(\eps)<0$ holds for all $\eps \in (0,{\overline{\varepsilon}_{DQN}}]$. Again, strong convexity of $\Phi$ and (\ref{s25}) imply that for all $\eps \in (0,{\overline{\varepsilon}_{DQN}}]$
{\small{
\begin{eqnarray*}
&\,&\Phi(x^{k+1}) - \Phi(x^*)  \leq (\Phi(x^k) \\
&\,&- \Phi(x^*))(1+{\mu_{\Phi}}\hat{\varphi}(\eps) )+e_{k} \leq (\Phi(x^k) - \Phi(x^*))\nu(\eps)+e_{k},
\end{eqnarray*}}}
where we recall ${\mu_{\Phi}}=\alpha \mu$, and where the last inequality comes from the fact that $\hat{\varphi}(\eps) \leq \varphi(\eps)$ for every $\eps>0$.

Finally, taking into account both cases and the fact that $e_{k}$ is nonnegative, we conclude that the following inequality holds with $\nu(\eps) \in (0,1)$ for all $\eps \in (0,{\overline{\varepsilon}_{DQN}}]$
$$\Phi(x^{k+1}) - \Phi(x^*)\leq (\Phi(x^k) - \Phi(x^*)) \nu(\eps) +e_{k}.$$
 $ \Box $

%
%First we prove that the iterations of the proposed algorithm are bounded in expectation provided that the step size $\varepsilon$ and the parameter $\rho$ are small enough.
%

\subsection{Mean square boundedness of the iterates and search directions}

We next show that
the iterates $x^k$ of DQN with idling are uniformly bounded
in the mean square sense.
Below, $E(\cdot)$ denotes the expectation operator.

\begin{lemma} \label{t1} Let the sequence of random variables
$\{x^k\}$ be generated by Algorithm~4, and let assumptions A1-A4
 hold. % Furthermore, let $\LL_{k}=-\rho_k \,\II$.
% and that $p_{k} \geq p_{min}$ for every $k \in \mathbb{N}$.
 Then, there exist
 %$p_{min} \in (0,1)$ and
  positive constants ${\overline{\rho}}$ and  ${\overline{\varepsilon}}$ depending on  $\alpha, \mu, L, \theta, \wmin, \wmax$ and $ p_{min}$, such that, for all $\rho \in [0,{\overline{\rho}}]$ and $\varepsilon \in (0, {\overline{\varepsilon}}]$,
  there holds: $E \left( \|x^k\|^2 \right) \leq \mathcal{C}_{\mathrm{x}}$, $k=0,1,...$,
  for some positive constant~$\mathcal{C}_{\mathrm{x}}$.
 \end{lemma}

{\em Proof.}
It suffices to prove that $E\left( \Phi(x^k)\right)$
 is uniformly bounded for all $k=0,1,...,$ since $\Phi$
  is strongly convex and therefore it holds that
 $\Phi(x) \geq \Phi(x^*)+\frac{{\mu_{\Phi}}}{2}\|x - x^*\|^2$, $x \in {\mathbb R}^{np}$. Further,
 for the sake of proving boundedness,
 without loss of generality we can assume that $f_{i}(x)\geq 0$, for all $x \in {\mathbb R}^p$, for every $i=1,...,n$.\footnote{Otherwise, since each of the $f_i$'s is lower bounded,
 we can re-define
 each $f_i$ as $\widehat{f}_i(x) = f_i(x) + c$,
 where $c$ is a constant larger than or
 equal $\max_{i=1,...,n}\left| \inf_{x \in {\mathbb R}^p} f_i(x)\right|$,
 and work with the $\widehat{f}_i$'s throughout the proof.}
 For $k=0,1,...,$ define
function $\Phi_k:\,{\mathbb R}^{n p} \rightarrow \mathbb R$,
by
\begin{eqnarray*}
&\,&\Phi_{k}(x)=\frac{\alpha}{p_{k}} F(x)+\frac{1}{2} x^{T}(\II-\ZZ)x=\frac{\alpha}{p_{k}} \sum_{i=1}^{n} f_{i}(x_{i})\\
&\,&+\frac{1}{2} {\sum_{\{i,j\}\in E,i<j}}w_{ij}\|x_{i}-x_{j}\|^2.
\end{eqnarray*}
Notice that $\Phi_{k}(x)=\Phi(x)$, $x \in {\mathbb R}^{np}$, if $p_{k}=1$.
Also note that,   for every $k =0,1,...$, we have:
$$\Phi(x) \leq \Phi_{k+1}(x) \leq \Phi_{k}(x)\leq \Phi_{0}(x),$$
since $p_{k}$ is assumed to be non-decreasing.
 The core of the proof is to
 upper bound $\Phi_{k+1}\left(x^{k+1}\right)$
 with a quantity involving $\Phi_{k}\left(x^{k}\right)$
 (see ahead~(34)), and after that to ``unwind''
 the resulting recursion.

To start, notice that $\Phi_{k}$ is strongly convex
and has Lipschitz continuous gradient for every $k$. More precisely, for every $k=0,1,...,$ and
for every
$x \in \mathbb{R}^{n p}$, we have that
$$\bar{\mu} \II \preceq \nabla^{2} \Phi_{k}(x) \preceq \bar{L} \II,$$
where $\bar{\mu}=\alpha \mu$ and $\bar{L}=\alpha L/ p_{min}+1$. Denote by
$$y_{k}=\nabla \Phi_{k}(x^{k})={\frac{\alpha}{p_{k}}}  \nabla F(x^{k})+(\II-\ZZ)\,x^{k}.$$
Further, let us define two more auxiliary maps,
as follows.
Let $\tik: \,{\mathbb R}^{np} \times \{0,1\}^{m} \rightarrow {\mathbb R}$,
and
$\bik: \,{\mathbb R}^{np} \times \{0,1\}^{m} \rightarrow {\mathbb R}$,
be given by:
\begin{eqnarray*}
\tik(x; \xi) &=& \frac{\alpha}{p_{k}} \sum_{i=1}^{n} \xi_i f_{i}(x_{i})+\frac{1}{2} {\sum_{\{i,j\}\in E,i<j}}w_{ij} \xi_i \xi_j \|x_{i}-x_{j}\|^2 \\
\bik(x; \xi) &=& \fik(x)-\tik(x)=\frac{\alpha}{p_{k}} \sum_{i=1}^{n}(1- \xi_i ) f_{i}(x_{i})\\
&+&\frac{1}{2} {\sum_{\{i,j\}\in E,i<j}}w_{ij}(1- \xi_i \xi_j) \|x_{i}-x_{j}\|^2,
\end{eqnarray*}
where $\{0,1\}^m$ denotes the set of all $m$-dimensional
vectors with the entries from set~$\{0,1\}$.
Introduce also the short-hand notation
\begin{eqnarray*}
{\widehat{\Phi}}_k(x) &=& \tik(x; \xi^k) = \frac{\alpha}{p_{k}} \sum_{i=1}^{n} \xik f_{i}(x_{i})\\
&\,&+\frac{1}{2} {\sum_{\{i,j\}\in E,i<j}}w_{ij} \xi_i^k \xi_j^k \|x_{i}-x_{j}\|^2 \\
{\widetilde{\Phi}}_k(x) &=& \bik(x; \xi^k) = \fik(x)-\tik(x)\\
&\,&=\frac{\alpha}{p_{k}} \sum_{i=1}^{n}(1- \xi_i^k ) f_{i}(x_{i})\\
&\,&+\frac{1}{2} {\sum_{\{i,j\}\in E,i<j}}w_{ij}(1- \xi_i^k \xi_j^k) \|x_{i}-x_{j}\|^2,
\end{eqnarray*}
where we recall that $\xi^k=(\xi_1^k,...,\xi_n^k)^T$
 is the node activation vector at iteration~$k$.
 Hence, note that,
 for any fixed~$x \in {\mathbb R}^{np}$,
 ${\widehat{\Phi}}_k(x)$ is
 a random variable, measurable
 with respect to the $\sigma$-algebra generated by~$\xi^k$.
  On the other hand,
  for any fixed
  value that variable~$\xi^k$ takes,
   $x \mapsto {\widehat{\Phi}}_k(x)$ is a
   (deterministic) function,
   mapping~${\mathbb R}^{np}$ to $\mathbb R$; analogous
  observations hold for $\hat{\Phi}_k$ as well.
   We will be interested
   in
   quantities ${\widehat{\Phi}}_k(x^k)$, ${\widetilde{\Phi}}_k(x^k)$
    ${\widehat{\Phi}}_k(x^{k+1})$, and ${\widetilde{\Phi}}_k(x^{k+1})$.
    Note that they are
    all random variables,
    measurable
     with respect to the
     $\sigma$-algebra
     generated by
     $\{\xi^s\}_{s=0,1,...,k}$.
     We
     will also work with the gradients
     of ${\widehat{\Phi}}_k$ and ${\widetilde{\Phi}}_k$ with respect to $x$, evaluated at $x^k$,
     that we denote by
$$\ty=\nabla \tik(x^k)=\frac{\alpha}{p_{k}} {{\mathbb{Y}_k}} \nabla F(x^{k})+(\II-\ZZ^{k}) x^{k}$$
and
$$\by=y_{k}-\ty=\nabla \bik(x^{k}).$$
These quantities
are also valid random variables,
measurable
     with respect to the
     $\sigma$-algebra
     generated by
     $\{\xi^s\}_{s=0,1,...,k}$.
 %Note that {\widehat{\Phi}}_k(x)

%
%
Now, recall that,
for each fixed $i$,
 $\xik$, $k=0,1,...,$ are independent identically distributed (i.i.d.) Bernoulli random variables.
  The same is true for the minimum and the maximum
$$\xim=\min_{i=1,...,n} \xik, \quad \ximax=\max_{i=1,...,n} \xik.$$
Also, notice that $E(\xim)=p_{k}^n$ and
\begin{eqnarray*}
&\,&\sqrt{1-\xim}=1-\xim, \,\,\, 1-\xik\leq 1-\xim,\\
&\,&1-\xik \xjk \leq 1-\xim.
\end{eqnarray*}

Consider now $\tik$ and $\ty$, regarded as functions of~$x \in {\mathbb R}^{n p}$. If $\ximax=1$ then $\tik$ is strongly convex
(with respect to $x $) with the same parameters as $\fik$, i.e.,
$$\bar{\mu} \II \preceq \nabla^{2} \tik(x) \preceq \bar{L} \II.$$
Now, denote by
${\hat{x}_{k}^{*}} = {\hat{x}_{k}^{*}}(\xi^k)$
the minimizer of $\tik$ with respect to~$x$;
using the fact that $\tik$ is nonnegative,
and that it has
a Lipschitz continuous gradient and is strongly convex, we obtain
$$\|\ty\|^2\leq \bar{L}^2 \|x^{k}-{\hat{x}_{k}^{*}}\|^2 \leq \frac{2\bar{L}^2 }{\bar{\mu}}(\tik(x^{k})-\tik({\hat{x}_{k}^{*}}))\leq  \frac{2\bar{L}^2 }{\bar{\mu}}\tik(x^{k}).$$
Using the fact that $\tik (x) \leq \fik(x)$, for all $x$, the previous inequality  yields
\be \label{prva}\|\ty\|\leq \bar{L} \sqrt{2/\bar{\mu}} \sqrt{\fik(x^k)}.\ee
On the other hand, if  $\ximax=0$ then ${{\mathbb{Y}_k}}=0$ and $\ZZ_{k}=\II$ which implies $\ty=0$ and the previous inequality obviously holds.

We perform a similar analysis considering $\bik$ and $\by$. First, notice that $\bik$ is also non-negative. Furthermore,  if $\xim=0$ then  $\bik$ is strongly convex with the same parameters as $\tik$ and the following holds
$$\|\by\|\leq \bar{L} \sqrt{2/\bar{\mu}} \sqrt{\bik(x^k)}.$$
Moreover, notice that
$$\bik(x^{k})\leq (1-\xim) \fik(x^{k})$$
and thus
\begin{eqnarray}
\label{druga}
&\,&\|\by\|\leq \bar{L} \sqrt{2/\bar{\mu}} \sqrt{(1-\xim)\fik(x^k)} \\
&\,&=(1-\xim)\bar{L} \sqrt{2/\bar{\mu}} \sqrt{\fik(x^k)}.
\end{eqnarray}

Let us return to $\fik$. This function also satisfies (\ref{21}), i.e.,
\be \label{21ap} \fik\left(x^{k}+\varepsilon\,s^k \right) \leq \fik(x^k)  + \frac{1}{2} \varepsilon^2 \bar{L} \|s^k\|^2 +  \varepsilon y_{k}^{T} s^k. \ee
 For the search direction~$\sk$  in step~2 of Algorithm~4,
and recalling ${\mathbb R}_k=(\II - \LL_k \GG^{k})\, {\mathbb A}_k^{-1}$
from Theorem~{IV.1}, we obtain that
$$\sk=-{\mathbb R}_k \ty.$$
Using the bounds (\ref{s2}) and (\ref{s1}) we conclude that $\|{\mathbb R}_k\|\leq (1+\rho {\mathcal{C}_{\mathrm{G}}}){\mathcal{C}_{\mathrm{A}}}:=\mathcal{C}_{\mathrm{R}}$ and therefore
\be \label{ap3} \|\sk\| \leq \mathcal{C}_{\mathrm{R}} \|\ty\|.\ee
 Also, by Theorem~{IV.1}, the following holds for
$\varepsilon \leq {\overline{\varepsilon}_{DQN}}$, and $\rho \leq
\overline{\rho}_{DQN}$ (where
$\overline{\rho}_{DQN}$ and ${\overline{\varepsilon}_{DQN}}$
are given in Theorems IV.1 and IV.2, respectively):
\be \label{ap4}\ty^{T}{\mathbb R}_k \ty =\ty^{T}\left( \frac{{\mathbb R}_k+{\mathbb R}_k^{T}}{2} \right) \ty\geq \delta \|\ty\|^{2}.\ee

From now on, we assume that $\varepsilon \leq {\overline{\varepsilon}_{DQN}}$, and $\rho \leq
\overline{\rho}_{DQN}$.
Substituting (\ref{ap3}) and (\ref{ap4}) into (\ref{21ap}) we obtain
\begin{eqnarray}
\fik (x^{k+1}) & \leq & \fik(x^k)  + \frac{1}{2} \varepsilon^2 \bar{L} R^2 \|\ty\|^2 -  \varepsilon (\ty+\by)^{T} {\mathbb R}_k \ty \nonumber\\
& \leq  &  \fik(x^k)  + \frac{1}{2} \varepsilon^2 \bar{L} R^2 \|\ty\|^2 \\
&-&  \varepsilon \delta \|\ty\|^2 +\varepsilon \mathcal{C}_{\mathrm{R}} \|\ty\|\|\by\| \nonumber \\
&=& \fik(x^k)  + (\frac{1}{2} \varepsilon^2 \bar{L} \mathcal{C}_{\mathrm{R}}^2 \nonumber \\
&-&  \varepsilon \delta) \|\ty\|^2 +\varepsilon \mathcal{C}_{\mathrm{R}} \|\ty\|\|\by\|  \label{as23}
 \end{eqnarray}
Since $(\frac{1}{2} \varepsilon^2 \bar{L} \mathcal{C}_{\mathrm{R}}^2 -  \varepsilon \delta)\leq 0$   for
$\eps \leq \frac{2 \delta}{\bar{L} \mathcal{C}_{\mathrm{R}}^2 }$, we conclude that
\begin{eqnarray*}
&\,&\fik (x^{k+1}) \leq  \fik(x^k)+\varepsilon \mathcal{C}_{\mathrm{R}} \|\ty\| \|\by\| \\
&\,&\leq
\fik(x^k)+\frac{2 \varepsilon  \bar{L} \mathcal{C}_{\mathrm{R}}^2}{\bar{\mu}} (1-\xim) \fik(x^{k}) ,
\end{eqnarray*}
for
\begin{equation}%************************
\label{eqn-eps-novo-appendix}
\eps \leq \min\left\{\frac{2 \delta}{\bar{L} \mathcal{C}_{\mathrm{R}}^2 }, {\overline{\varepsilon}_{DQN}}\right\},
\,\,\rho \leq \overline{\rho}_{DQN},
\end{equation}
where ${\overline{\varepsilon}_{DQN}}$ and $\overline{\rho}_{DQN}$ are given in Theorems IV.1 and IV.2.
Denoting $B=\frac{2 \varepsilon \bar{L} \mathcal{C}_{\mathrm{R}}^2}{\bar{\mu}}$, and using the fact that
$\Phi_{k+1}(x) \leq \fik (x)$,
for all $x \in {\mathbb R}^{n p}$, we obtain
\be \label{ap5} \Phi_{k+1} (x^{k+1}) \leq (1+B(1-\xim)) \fik(x^{k}) .\ee
Applying expectation we obtain
$$E(\Phi_{k+1} (x^{k+1})) \leq (1+B(1-p_{k}^n)) E(\fik(x^{k})),$$
where we use independence between $\xim$ and $x^k$.
Furthermore, recall that $u_{k}=1-p_{k}$ and notice that $1-p_{k}^{n}\leq n u_{k}$. Moreover, $1+t\leq e^{t}$ for $t>0$ and thus
$$E(\Phi_{k+1} (x^{k+1})) \leq e^{nB u_{k}} E(\fik(x^{k})).$$
Next, by unwinding the recursion, we obtain
$$E(\fik (x^{k})) \leq e^{nB \sum_{j=0}^{k-1} u_{j}} \Phi_{0}(x^{0}):= {\mathcal{C}_{\Phi,2}}.$$
By assumption, $\{u_{k}\}$ is summable, and since $\Phi(x)\leq \fik(x)$,
for all $x$, we conclude that
$$E(\Phi (x^{k})) \leq {\mathcal{C}_{\Phi,2}} .$$
Finally, since $\Phi$ is strongly convex, the desired result holds.
$\Box$

%*********
%A couple of remarks are now in order.

%{\em Remark~1.} We can prove boundedness
%of iterates~$x^k$ also for
%the generic ${\mathbb L}_k$'s that
%satisfy the safeguarding condition
%(condition~(1) in Algorithm~4,
%for any $\rho \in [0,{\overline{\rho}}^\prime]$,
%for some positive ${\overline{\rho}}^\prime$),
%under the additional assumption
%that $p_k=1-u_k$,
%and that the sequence $\{u_k\}$
% is summable. The proof of the result
% under the alternative set of conditions
% is provided in the Appendix.
% This translates into requiring that $p_k$ to converges to one
% sufficiently fast. For example,
% it suffices to
% have $u_k=\frac{1}{(k+1)^{1+\zeta}}$,
% $\zeta>0$ arbitrarily small.
%

%{\em Remark~2}. The set of admissible
%parameters~$\epsilon,\rho$ in Lemma~\ref{t1}, see equations \eqref{eqn-eps-novo} and \eqref{eqn-rho-novo},
% for which the boundedness
%of the iterates
%is ensured is more
%conservative  than
%with the
%convergence analysis of
%the standard DQN in~\cite{DQN}.
%This is the price paid by
%allowing for very generic sequences
%$\{p_k\}$. In the Appendix,
%we show that
%assuming a summable sequence
%$\{u_k\}$ enables to significantly
%widen the range of admissible
%parameters $\epsilon,\rho$ (see ahead
%\eqref{eqn-eps-novo-appendix} in the Appendix).
%%,
%%practically matching the case of DQN.
%%[***PROVERITI IPAK POSLEDNJU IZAVU.]
%
%

Notice that an immediate consequence of Lemma~{IV.1} is that the gradients are uniformly bounded in the mean square sense. Indeed,
\begin{eqnarray}
E\left(\| \nabla F (x^{k}) \|^2\right) & =  & E\left(\| \nabla F (x^{k})-\nabla F (\tilde{x}^{*}) \|^2\right) \nonumber\\
& \leq  & E\left(L^2 \| x^{k}-\tilde{x}^{*} \|^2\right) \nonumber\\
&\leq & 2 L^2 (E\left(\| x^{k} \|^2\right)+\| \tilde{x}^{*} \|^2)\nonumber\\
& \leq & 2 L^2 ({\mathcal{C}_{\mathrm{x}}}+\| \tilde{x}^{*} \|^2):={\mathcal{C}_{\mathrm{F}}},  \label{s8}
\end{eqnarray}
where we recall $\mathcal{C}_{\mathrm{x}}$ in Lemma~\ref{t1}.

Next, we show that the ``inexactness''
of the search directions of DQN with idling
are ``controlled'' by the activation probabilities~$p_k$'s.

\begin{te} \label{t2} Let assumptions A1-A4 hold, and consider ${\overline{\rho}}$ and  ${\overline{\varepsilon}}$ as in Lemma~\ref{t1}.
 %Assume that $\LL_{k}=-\rho_k \II$.
Then, for all $\rho \in [0,{\overline{\rho}}]$ and $\varepsilon \in (0, {\overline{\varepsilon}}]$,
  the following inequality  holds for every $k$  and some positive constant ${\mathcal{C}_{\mathrm{s}}}$:
\begin{equation}
\label{eqn-C-s}
E(\|\sk-\skt\|^2)\leq (1-p_{k}) {\mathcal{C}_{\mathrm{s}}}.
\end{equation}
 \end{te}
{\em Proof.}

We first split the error as follows:
{\allowdisplaybreaks{\begin{eqnarray}
&\,&E\left(\| \sk-\skt \|^2\right)  \nonumber\\
&\,&\,\,\,=   E\left(\| \hk \gk-\hkt \gkt+\hkt \gk- \hkt \gk\|^2\right) \nonumber\\
& \,& \,\,\,\leq   2 \left(E\left( \| \hkt\|^2 \|(\gk-\gkt) \|^2\right)+E\left( \| (\hk-\hkt) \gk \|^2\right)\right) \nonumber\\
&\,&\,\,\,\leq  2 \left(({\mathcal{C}_{\mathrm{H}}})^2 E\left( \|(\gkt-\gk) \|^2\right) \right. \nonumber\\
&\,& \,\,\,\left. +E\left( \| (\hk-\hkt) \gk \|^2\right)\right).  \label{s9}
\end{eqnarray}}}
We will estimate the expectations above separately.
 Start by observing that:
\begin{eqnarray}
&\,&\gkt_{i}-\gk_{i} =  -(A^{k}_{i})^{-1} \left(\alpha \nabla f_{i} (x_{i}^{k})+\sum_{j\in O_{i}}w_{ij}(x^{k}_{i}-x^{k}_{j})\right)  \nonumber\\
& \,&\,\,\,\,\,+\,\,   \xik (A^{k}_{i})^{-1} \left(\frac{\alpha}{p_{k}} \nabla f_{i} (x_{i}^{k})+\sum_{j\in O_{i}}w_{ij} \xjk (x^{k}_{i}-x^{k}_{j})\right) \nonumber\\
& \,&\,\,\,\,\,=\,\,  (\frac{\xik}{p_{k}}-1) \alpha (A^{k}_{i})^{-1}\nabla f_{i} (x_{i}^{k}) \nonumber \\
&\,&\,\,\,\,\,+\,\,
 (A^{k}_{i})^{-1} \sum_{j\in O_{i}}w_{ij} (\xik \xjk-1) (x^{k}_{i}-x^{k}_{j}).
 \nonumber
\end{eqnarray}
Notice that  (\ref{s2}) implies
$\|(\frac{\xik}{p_{k}}-1) \alpha (A^{k}_{i})^{-1}\nabla f_{i} (x_{i}^{k})\|^2 \leq 2 (\frac{\xik}{p_{k}}-1)^2 \alpha^2 ({\mathcal{C}_{\mathrm{A}}})^2 \|\nabla f_{i} (x_{i}^{k})\|^2$,
 with $\mathcal{C}_{\mathrm{A}}$ defined in \eqref{s2}. Also, the assumptions on the $w_{ij}$'s (Assumption~{A3}) and convexity of the scalar
quadratic function $\mathcal{V}(\mu) =  \mu^2$ yield:
\begin{eqnarray}
&\,&\left\|\sum_{j\in O_{i}}w_{ij} (\xik \xjk-1) (x^{k}_{i}-x^{k}_{j})\right\|^2 \nonumber \\
& \,&\,\,\,\,\,\,\,\,\,\,\,\,\,\,\,\,\,\,\leq \,\, \left(\sum_{j\in O_{i}}w_{ij} (1-\xik \xjk) \|x^{k}_{i}-x^{k}_{j}\|\right)^2 \nonumber\\
& \,&\,\,\,\,\,\,\,\,\,\,\,\,\,\,\,\,\,\,\leq \,\, \sum_{j\in O_{i}}w_{ij} (1-\xik \xjk)^2 \|x^{k}_{i}-x^{k}_{j}\|^2. \nonumber
 \end{eqnarray}
Therefore,
\begin{eqnarray*}
&\,&\| \gkt_{i}-\gk_{i}\|^2\leq 2 (\frac{\xik}{p_{k}}-1)^2 \alpha^2 ({\mathcal{C}_{\mathrm{A}}})^2 \|\nabla f_{i} (x_{i}^{k})\|^2 \\
&\,&  +  2({\mathcal{C}_{\mathrm{A}}})^2 \sum_{j\in O_{i}}w_{ij} (1-\xik \xjk)^2 \|x^{k}_{i}-x^{k}_{j}\|^2.
  \end{eqnarray*}
Applying the expectation and using the fact that $\xik$'s are
independent, identically distributed (i.i.d.) across $i$ and across iterations, the fact that
\[E((\frac{\xik}{p_{k}}-1)^2)=(1-p_{k})/p_{k} \leq (1-p_{k})/p_{min},\] and
\[E((1-\xik \xjk)^2)=1-p_{k}^2=(1-p_{k})(1+p_{k}) \leq 2(1-p_{k})\] we obtain the following inequality
\begin{eqnarray*}
&\,&E(\| \gkt_{i}-\gk_{i}\|^2) \leq 2 ({\mathcal{C}_{\mathrm{A}}})^2 (1-p_{k}) (\frac{\alpha^2}{p_{min}}E(\|\nabla f_{i} (x_{i}^{k})\|^2) \\
&\,&+
2\sum_{j\in O_{i}}w_{ij}  E(\|x^{k}_{i}-x^{k}_{j}\|^2)).
\end{eqnarray*}
Further, note that
$E(\| \gkt-\gk\|^2) = \sum_{i=1}^{n} E(\| \gkt_{i}-\gk_{i}\|^2) $. Moreover, as a consequence of Lemma \ref{t1} we have
$\sum_{i=1}^{n} E(\|\nabla f_{i} (x_{i}^{k})\|^2)=E(\|\nabla F(x^{k})\|^2) \leq {\mathcal{C}_{\mathrm{F}}}$, and
{\allowdisplaybreaks{
\begin{eqnarray}
&\,&\sum_{i=1}^{n} \sum_{j\in O_{i}}w_{ij} E(\|x^{k}_{i}-x^{k}_{j}\|^2) \nonumber \\
&\leq&
\sum_{i=1}^{n} \sum_{j\in O_{i}}w_{ij} \,2 \,E(\|x^{k}_{i}\|^2+\|x^{k}_{j}\|^2) \nonumber\\
& = & 2(E(\sum_{i=1}^{n} \|x^{k}_{i}\|^2 \sum_{j\in O_{i}}w_{ij})+
E( \sum_{j\in O_{i}} \|x^{k}_{j}\|^2 \sum_{i=1}^{n}w_{ij})) \nonumber\\
& \leq  & 2(E(\|x^{k}\|^2 +E(\|x^{k}\|^2) )
\nonumber\\
 &\leq & 4 {\mathcal{C}_{\mathrm{x}}}, \label{s14}
 \end{eqnarray}}}
 where $\mathcal{C}_{\mathrm{x}}$ is given in Lemma~\ref{t1}.
Combining the bounds above, we conclude
\be \label{s12} E(\| \gkt-\gk\|^2) \leq  (1-p_{k}) 2 ({\mathcal{C}_{\mathrm{A}}})^2 (\frac{\alpha^2}{p_{min}}{\mathcal{C}_{\mathrm{F}}}+8 {\mathcal{C}_{\mathrm{x}}}):=(1-p_{k}) {\mathcal{C}_{\mathrm{g}}}.\ee
Now, we will estimate the second expectation term in (\ref{s9}). Again, consider an arbitrary block $i$ of the vector under expectation. We obtain the following:
\begin{eqnarray}
[(\hk \hspace{-3mm}&-& \hspace{-3mm}\hkt) \gk]_{i}  = [\LL_{k} (\GG-\GG^k) \gk]_{i} \nonumber \\
&=& [\LL_{k} (\ZZ_{u}-\ZZ_{u}^{k}+\theta (\ZZ_{d}^{k}-\ZZ_{d})) \gk]_{i} \nonumber\\
& = & \Lambda_{i}^{k} \sum_{j \in O_{i}} (w_{ij}-w_{ij}^{k})\gk_{j}+ \Lambda_{i}^{k} \theta (w^{k}_{ii}-w_{ii})\gk_{i} \nonumber\\
& =  & \Lambda_{i}^{k} \sum_{j \in O_{i}} w_{ij}(1-\xik \xjk)\gk_{j}+ \Lambda_{i}^{k} \theta \sum_{j \in O_{i}} (w_{ij}-w_{ij}^{k})\gk_{i}
\nonumber\\
 &= & \Lambda_{i}^{k} \sum_{j \in O_{i}} w_{ij}(1-\xik \xjk)(\gk_{j}+\theta \gk_{i}). \nonumber
 \end{eqnarray}
 Moreover, applying the norm and the convexity argument like above we get
 \begin{eqnarray}
\|[(\hk \hspace{-3mm} &-&  \hspace{-3mm}\hkt) \gk]_{i}\|^2 \leq \|\Lambda_{i}^{k}\|^2 \|\sum_{j \in O_{i}} w_{ij}(1-\xik \xjk)(\gk_{j}+\theta \gk_{i})\|^2 \nonumber\\
& \leq  & \rho^2 \sum_{j \in O_{i}} w_{ij}(1-\xik \xjk)^2 \|\gk_{j}+\theta \gk_{i}\|^2. \label{s15}
 \end{eqnarray}
 Therefore, $E(\|[(\hk-\hkt) \gk]_{i}\|^2)\leq \rho^2 \sum_{j \in O_{i}} w_{ij}2(1-p_{k}) E \|\gk_{j}+\theta \gk_{i}\|^2)$
and using the steps similar to the ones in (\ref{s14}) we obtain the inequality
$$E(\|(\hk-\hkt) \gk \|^2) \leq (1-p_{k}) \rho^2 4 (1+\theta^2) E(\|\gk\|^2).$$
 Moreover,
 \begin{eqnarray}
E(\|\gk\|^2) & = & E(\|{\mathbb A}_k^{-1} ({\frac{\alpha}{p_{k}}} {{{\mathbb{Y}_k}}} \nabla F(x^{k})+(\II-\ZZ^{k})x^{k})\|^2)  \nonumber\\
& \leq  & ( {\mathcal{C}_{\mathrm{A}}})^2 2(\frac{\alpha^2}{p_{min}^2}E(\|{{{\mathbb{Y}_k}}}\|^2) E( \|\nabla F(x^{k})\|^2) \nonumber \\
 &+& E(\|\II-\ZZ^{k}\|^2) E( \|x^{k}\|^2))\,) \nonumber\\
& \leq & ({\mathcal{C}_{\mathrm{A}}})^2 2(\frac{\alpha^2}{p_{min}^2} {\mathcal{C}_{\mathrm{F}}}+ {\mathcal{C}_{\mathrm{x}}}):={\mathcal{C}_{\mathrm{g,2}}}\label{s17}
 \end{eqnarray}
 and thus $$E(\|(\hk-\hkt) \gk \|^2) \leq (1-p_{k}) \rho^2 4 (1+\theta^2) {\mathcal{C}_{\mathrm{g,2}}}.$$
 Finally, returning to (\ref{s9}), the previous inequality and (\ref{s12}) imply
 $$ E\left(\| \sk-\skt \|^2\right) \leq (1-p_{k}) 2 (({\mathcal{C}_{\mathrm{H}}})^2{\mathcal{C}_{\mathrm{g}}}+\rho^{2} {\mathcal{C}_{\mathrm{g,2}}}):= (1-p_{k}) {\mathcal{C}_{\mathrm{s}}} .$$
$ \Box $

\subsection{Main results}
The next result -- first main result -- follows from the theorems stated above. Let us define constants ${{\overline{\rho}}_{idl}}= \min \{{\overline{\rho}},{\overline{\rho}}_{DQN}\}$ and $ {{\overline{\varepsilon}}_{idl}}=\min \{{\overline{\varepsilon}}, {\overline{\varepsilon}_{DQN}}\}$ where ${\overline{\rho}}$ and ${\overline{\varepsilon}}$ are as in Theorem \ref{t2} and ${\overline{\rho}}_{DQN}$ and ${\overline{\varepsilon}_{DQN}}$ are as in Theorem \ref{t3}.

 \begin{te} \label{t4}
 Let $\{x^k\}$ be the sequence
 of random variables generated by Algorithm~4.
 Further, let assumptions A1-A4 hold.
  In addition, let $\rho \in [0, {{\overline{\rho}}_{idl}}]$ and
$\eps \in (0, {{\overline{\varepsilon}}_{idl}}]$.
Then:
   \be \label{31} E \left(\Phi(x^{k+1}) - \Phi(x^*)\right) \leq E \left(\Phi(x^k) - \Phi(x^*)\right) \nu +{\mathcal{C}_{\Phi}} (1-p_{k}),\ee
   where $\nu \in (0,1)$ is a constant, and ${\mathcal{C}_{\Phi}}$ is a positive constant.
   %${\mathcal{C}_{\Phi}}={\mathcal{C}_{\mathrm{s}}} (\eps^2 {L_{\Phi}}+\eps / q) $ is a positive constant.
    Moreover,
    the iterate sequence $\{x^k\}$
      converges to
       the solution~$x^{*}$ of~\eqref{reformulation1}
       in the mean square sense and almost surely.
       \end{te}

{\em Proof}.
Claim~\eqref{31}
follows by taking expectation in Theorem~{IV.3}.
 The remaining two claims
 follow similarly to
 the proof of Theorem~2 in~\cite{DGDSN}.
  We briefly demonstrate the
  main arguments for completeness.
  Namely,
  unwinding
 the recursion~\eqref{31},
 we obtain for $k=1,2,...$:
 \begin{eqnarray}
 \label{eqn-31-1}
E (\Phi(x^{k}) \hspace{-3mm}&-& \hspace{-3mm}\Phi(x^*))
 \leq
 \left(\Phi(x^{0}) - \Phi(x^*)\right)
 \nu^{k}\\
 &\,&\,\,\,\,\,\,\,\,\,\,\,\,\,\,\,\,\,+\,\,{\mathcal{C}_{\Phi}}\,\sum_{t=1}^k
 \nu^{k-t}(1-p_{t-1}). \nonumber
 \end{eqnarray}
Now, we apply Lemma~{IV.1}.
%3.1 in \cite{RamNedicVeeravalli}
% which says that, for
% a (deterministic) sequence
%  $\{a_k\}$ converging to zero,
%  and $\nu \in (0,1)$, it holds that:
%  \[
%  \sum_{t=1}^k
% \nu^{k-t}a_{t-1} \rightarrow 0\mathrm{\,\,as\,\,}k \rightarrow \infty.
%  \]
  From this result and~\eqref{eqn-31-1},
 it follows directly that
  $E \left(\Phi(x^{k}) - \Phi(x^*)\right)
   \rightarrow 0$ as $k \rightarrow \infty$,
   because it is assumed that $p_k \rightarrow 1$.
    Furthermore,
    using
    inequality
     $\Phi(x^{k}) - \Phi(x^*)
     \geq \frac{{\mu_{\Phi}}}{2} \|x^k-x^*\|^2$,
     the mean square convergence
     of $x^k$ towards $x^*$ follows.
      It remains to show that
      $x^k \rightarrow x^*$
      almost surely, as well.
      Using condition~\eqref{eqn-u-k-cond},
       inequality~\eqref{eqn-31-1}
       implies that:
 \begin{eqnarray}
 \label{eqn-31-2}
E \left(\Phi(x^{k}) - \Phi(x^*)\right)
 &\leq&
 \left(\Phi(x^{0}) - \Phi(x^*)\right)
 \nu^{k}\\
 &+&{\mathcal{C}_{\Phi}}\,{\mathcal{C}_{\mathrm{u}}}\,\sum_{t=1}^k
 \frac{\nu^{k-t}}{t^{1+\zeta}}. \nonumber
 \end{eqnarray}
   It can be shown
   that~\eqref{eqn-31-2}
    implies (see, e.g.,~\cite{DGDSN}):
     $E \left(\Phi(x^{k}) - \Phi(x^*)\right)=O\left(1/k^{1+\zeta}\right)$,
      which further implies:
      \begin{equation}
      \label{eqn-31-3}
      E \left(\|x^k - x^*\|^2\right)=O\left(1/k^{1+\zeta}\right).
      \end{equation}
    Applying the Markov inequality
    for the random variable $\|x^k-x^*\|^2$,
    we obtain, for any~$\kappa >0$:
    \begin{equation}
      \label{eqn-31-4}
    P \left(\|x^k - x^*\|^2 > \kappa\right)\leq \frac{1}{\kappa} E \left(\|x^k - x^*\|^2\right)=  O\left(1/k^{1+\zeta}\right).
   \end{equation}
   Inequality~\eqref{eqn-31-4}
   implies that:
   \[
   \sum_{k=0}^{\infty} P \left(\|x^k - x^*\|^2 > \kappa\right) < \infty,
   \]
   and so, by the first Borel-Cantelli lemma, we get
    $
    P \left(\|x^k - x^*\|^2 > \kappa, \,\,\mathrm{infinitely\,\,often}\right)
     = 0,
    $
     which implies that $x^k \rightarrow x^*$, almost surely. $\Box$

Next, we state and prove our second main result.
\begin{te}
\label{t5}
Let $\{x^k\}$ be the sequence
 of random variables generated by Algorithm~4. Further,
 let assumptions of Theorem~\ref{t4} hold, and let~$p_{k}=1-\sigma^{k+1}$, with
$\sigma \in (0,1)$.
Then, $\{x^k\}$ converges to the solution~$x^{*}$ of problem~(\ref{reformulation1})
    in the mean square sense at an R-linear rate.
%the convergence in a mean square sense is R-linear, i.e.
%$E(\|x^{k}-x^{*}\|^2)$ tends to zero R-linearly.
 \end{te}
 {\em Proof.}
Denote $z_{k}={\mathcal{C}_{\Phi}} (1-p_{k})$.  For the specific choice of $p_{k}$ we obtain $z_{k}={\mathcal{C}_{\Phi}} \sigma^{k+1}$ which obviously converges to zero R-linearly. Furthermore, repeatedly applying the relation  (\ref{31}) we obtain
$$E \left(\Phi(x^{k}) - \Phi(x^*)\right) \leq (\Phi(x^0) - \Phi(x^*)) \nu^{k} + a_{k} $$
where $a_{k}=\sum_{j=1}^{k} \nu^{j-1} z_{k-j}.$
%Since $z_{k}$ tends to zero R-linearly and $\nu \in (0,1)$,
Moreover, it can be shown (see Lemma~{II.1}) that $a_{k}$ also converges to zero R-linearly which implies the R-linear convergence of $E \left(\Phi(x^{k}) - \Phi(x^*)\right)$.
Now, using the strong convexity of $\Phi$ (with strong convexity constant ${\mu_{\Phi}}=\alpha \mu$),  we get: $\|x^{k}-x^{*} \|^2 \leq (\Phi(x^{k}) - \Phi(x^*)) 2/ {\mu_{\Phi}}$, which in turn implies:
$$E \left(\|x^{k}-x^{*} \|^2 \right) \leq E \left(\Phi(x^{k}) - \Phi(x^*)\right) 2/ {\mu_{\Phi}}.$$
The last inequality means that $E \left(\|x^{k}-x^{*} \|^2 \right) $ also converges to zero R-linearly.
%, i.e. there exist constants $\varrho \in (0,1)$ and $V>0$ such that for every $k$
% \be \label{32} E \left(\|x^{k}-x^{*} \|^2 \right) \leq V \varrho^k .\ee
 %Moreover, Theorem \ref{t1} ensures that $E \left(\|x^{k} \|^2 \right) $ is uniformly bounded by the constant ${\mathcal{C}_{\mathrm{x}}}$ and therefore we conclude that $x^{k}$ converges to $x^{*}$ in a mean square sense.
%
%In order to prove the almost sure convergence, let us take an arbitrary $\epsilon>0$. Chebyshev's inequality  yields
%$$P \left(\|x^{k}-x^{*} \| \geq \epsilon \right) \leq E \left(\|x^{k}-x^{*} \|^2 \right)/\epsilon^2$$
%and combining this with (\ref{32}) we obtain
%$$\sum_{k \in \mathbb{N}} P \left(\|x^{k}-x^{*} \| \geq \epsilon \right) \leq \frac{V}{\epsilon} \sum_{k \in \mathbb{N}} \varrho^k < \infty.$$
%Finally, using the argument of the Borel-Cantelli lemma, we conclude that   $x^{k}$ converges to $x^{*}$ almost surely.
  $ \Box $

Theorem~IV.5
shows that the DQN method with idling
converges at an R-linear rate when $p_k=1-\sigma^{k+1}$.
 Parameter~$\sigma$ plays an important role
 in the practical performance of the method.
We recommend the tuning
 $\sigma=1-c\,\alpha\,\mu$,
 with $c \in [30,80]$, for $\alpha < 1/(80 L)$. \footnote{Condition~$\alpha < 1/(80 L)$
  is not restrictive, as
  we observed experimentally that
  usually one needs to take
  an $\alpha$ smaller
  than $1/(80 L)$ in order to achieve a
  satisfactory limiting accuracy.}
  The rationale for the tuning above
   comes from distributed first order methods
   with idling~\cite{DGDSN},
   where we showed analytically that
   it is optimal (in an appropriate sense)
    to set~$\sigma^{\bullet}=(1-\alpha \mu)^2 \approx 1-2 \alpha \mu$.
     The value~$\sigma^{\bullet}$
     is set to balance 1) the
     linear convergence factor of the method without idling
     and 2) the convergence factor of the convergence of $p_k$ to one.
     As DQN has a better (smaller) convergence factor than
     the distributed first order method (due
     to incorporation of the second order information),
     one has to adjust the rule~$1-2 \alpha \mu$,
     replacing $2$ with a larger constant~$c$;
     experimental studies suggest values for $c$ on the order~$30$-$80$.
      In addition, for very small values of~$\alpha \mu$,
      in order to prevent very small $p_k$'s at initial iterations on the one hand
      and the $\sigma$'s very close to one on the other hand,
      we can utilize a ``safeguarding'' and modify $p_k$ to
      $p_k=\max\left\{ \underline{p},1-(\min\{\sigma,\overline{\sigma}\})^{k+1}\right\}$,
      where $\underline{p}$
       can be taken, e.g., as~$0.2$, and
       $\overline{\sigma}$ as~$0.9999.$

\section{Extensions: DQN under persisting idling}

%We have just proved that controlled idling does not effect convergence too much. Indeed, the sequence of iterations converge to the solution almost surely and in a mean square sense, while the later is with R-linear rate. One of the necessary conditions for the convergence is that the probability of activation converges to 1 fast enough.

%However, there are some situations where this probability can not be fully controlled. For example, there can be some external factors that affect the activity of the nodes while their influence is out of our control.

This section investigates DQN with idling
when activation probability~$p_k$
 does not converge to one asymptotically,
 i.e., the algorithm is subject to
 persisting idling.
 This scenario is of interest when
 activation probability~$p_k$
  is not in full control of
  the algorithm designer
   (and the networked nodes during execution).
     For example, in applications
like wireless sensor networks, inter-node
messages may be lost, due to, e.g.,
random packet dropouts. In addition, an active node may fail
to perform its solution estimate update at a certain iteration,
because the actual
calculation may take longer than the time slot allocated for
one iteration, or simply due to unavailability of sufficient computational
resources.

Henceforth, consider the scenario when
$p_k$ may not converge to one. In other words,
regarding Assumption A4, we only keep the
requirement that the sequence $\{p_k\}$
 is uniformly bounded from below.
 We make here an additional assumption
 that the iterates are bounded
 in the mean square sense,
 i.e., $E\left(\|x^k\|^2\right)$
 is uniformly bounded from above by
 a positive constant. Now, consider relation (\ref{31});  it continues to hold under the assumptions of the Theorem~\ref{t4}, i.e., we have
 \begin{eqnarray*}
 E \left(\Phi(x^{k}) - \Phi(x^*)\right) &\leq& (\Phi(x^0) - \Phi(x^*)) \nu^{k} \\
 &+& \sum_{j=1}^{k} \nu^{j-1} {\mathcal{C}_{\Phi}} (1-p_{k-j}).
 \end{eqnarray*}
 Therefore, using the previous inequality we obtain
$$E \left(\Phi(x^{k}) - \Phi(x^*)\right) \leq (\Phi(x^0) - \Phi(x^*)) \nu^{k} + \frac{{\mathcal{C}_{\Phi}} (1-p_{min})}{1-\nu}.$$
Using strong convexity of $\Phi$ (with strong convexity constant ${\mu_{\Phi}}$) and letting $k$ go to infinity we obtain
$$\limsup_{k \rightarrow \infty} E \left(\|x^{k}-x^{*} \|^2 \right) \leq  \frac{2 {\mathcal{C}_{\Phi}} (1-p_{min})}{{\mu_{\Phi}}(1-\nu)}:=\mathcal{E}. $$
Therefore, the proposed algorithm converges
(in the mean square sense) to a neighborhood of
the solution~$x^*$ of (2).
Hence, an additional limiting error
(in addition to the error due to the difference between
the solutions of (1) and (2)) is introduced
with respect to the case $p_k \rightarrow 1$.
In order to analyze further quantity~$\mathcal{E}$, we unfold ${\mathcal{C}_{\Phi}}$ to get
$$\mathcal{E}=(1-p_{min}) h(\eps) l(\rho),$$
where $$h(\eps)=\frac{\eps^2 {L_{\Phi}}+\eps/q}{1-\nu(\eps)},\,\,\, l(\rho)=\frac{4}{{\mu_{\Phi}}} ((1+\rho {\mathcal{C}_{\mathrm{G}}})^2 {\mathcal{C}_{\mathrm{g}}}+\rho^2 {\mathcal{C}_{\mathrm{g,2}}});$$
and $\nu(\eps)$ is as in the proof of Theorem \ref{t3}.
(Recall ${\mathcal{C}_{\mathrm{G}}}$,
${\mathcal{C}_{\mathrm{g}}}$, and ${\mathcal{C}_{\mathrm{g,2}}}$
 in \eqref{s1}, \eqref{s12}, and \eqref{s17}, respectively.)
 %i.e. $\nu(\eps)=1+{\mu_{\Phi}} \varphi(\eps)$ and $\varphi(\eps)=\varepsilon^2 {L_{\Phi}}(\beta^2+q^2)-\eps (\delta-q)$.
It can be shown that $h(\eps)$ is an increasing function of $\eps$ so taking smaller step size $\eps$ brings us closer to the solution. However, the convergence factor $\nu(\eps)$ is also increasing with  $\eps$; thus, there is a tradeoff between the precision and the convergence rate. Furthermore, considering $l(\rho)$, we can see that it is also an increasing function. However, as it is expected, $l(0)$ is strictly positive. In other words, the error remains positive
when the safeguarding parameter $\rho=0$.
Finally, the size of the error is proportional to $(1-p_{min})$ --
the closer $p_{min}$ to one, the smaller the error.
 Simulation examples in Section~6
 demonstrate that the error
 is only moderately increased
 (with respect to the case $p_k \rightarrow 1$),
 even in the presence of very strong persisting idling.

\section{Numerical results}

This section demonstrates by simulation significant computational and communication savings
 incurred through the idling mechanism within DQN.
 It also shows that persisting idling ($p_k$ not converging to one)
  induces only a moderate additional limiting error, i.e.,
  the method continues to converge to a solution neighborhood
  even under persisting idling.

\begin{figure}[thpb]
      \centering
      \subfigure[$\,$]{\includegraphics[height=2.6 in,width=3.2 in]{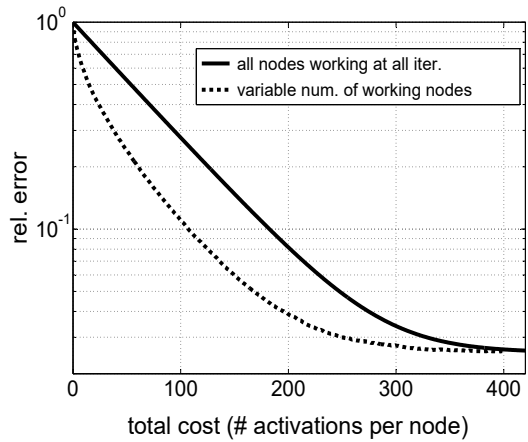}}
      \subfigure[$\,$]{\includegraphics[height=2.6 in,width=3.2 in]{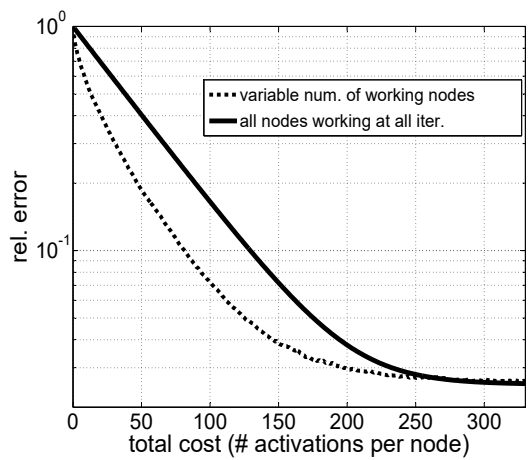}}
      \caption{Relative error versus total cost (number of activations per node) for
      quadratic costs and $n=100$-node network; Figure~(a): $\Lambda_i^k=0$; Figure~(b): $\Lambda_i^k=-I$.
      The solid lines correspond to all nodes working at all iterations;
      the dashed lines correspond to the method with idling (increasing number of working nodes).}
      \label{Figure_quadratic_small}
\end{figure}

\begin{figure}[thpb]
      \centering
      \subfigure[$\,$]{\includegraphics[height=2.6 in,width=3.2 in]{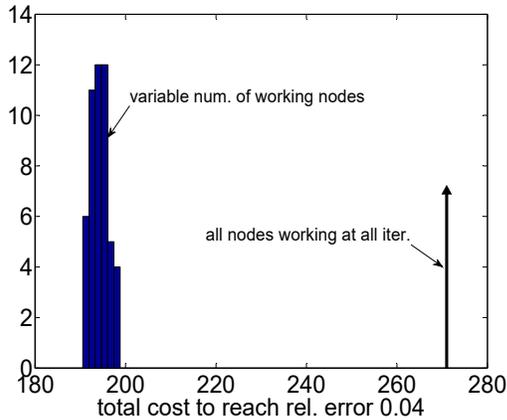}}
      \subfigure[$\,$]{\includegraphics[height=2.6 in,width=3.2 in]{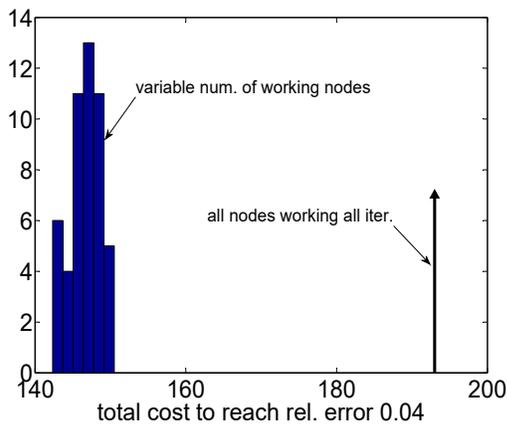}}
      \caption{Total cost (number of activations per node) to
      reach relative error~$0.04$ for
      quadratic costs and $n=100$-node network; Figure~(a): $\Lambda_i^k=0$; Figure~(b): $\Lambda_i^k=-I$.
      The histograms corresponds to the DQN algorithm with idling;
      the arrow indicates the total cost needed by standard DQN.}
      \label{Figure_quadratic_small}
\end{figure}

\begin{figure}
    \centering
    \subfigure[$\,$]{\includegraphics[height=2.6 in,width=3.2 in]{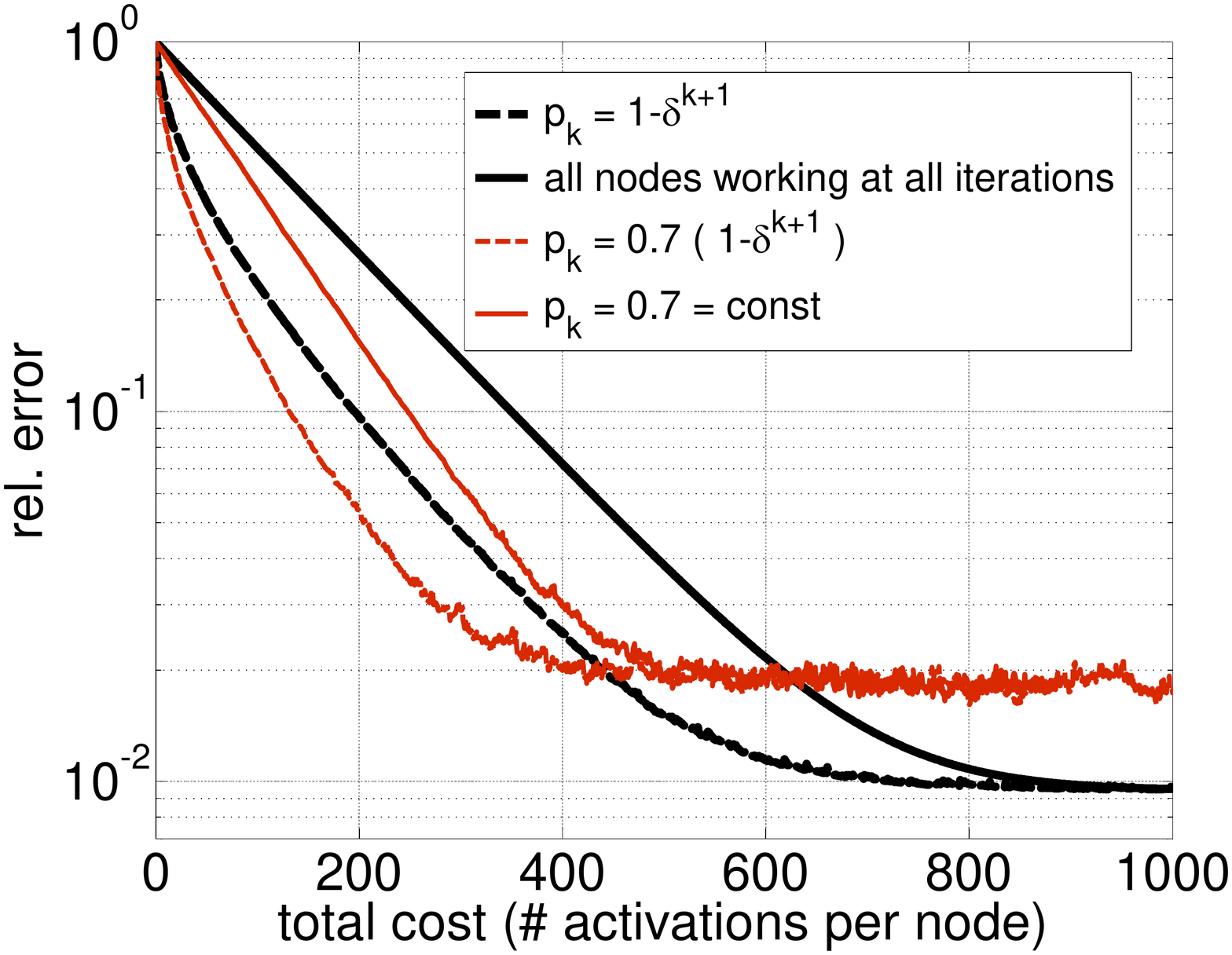}}
    %{gull}
        %\caption{A gull}
        %\label{fig:gull}
    ~ %add desired spacing between images, e. g. ~, \quad, \qquad, \hfill etc.
   \subfigure[$\,$]{\includegraphics[height=2.6 in,width=3.2 in]{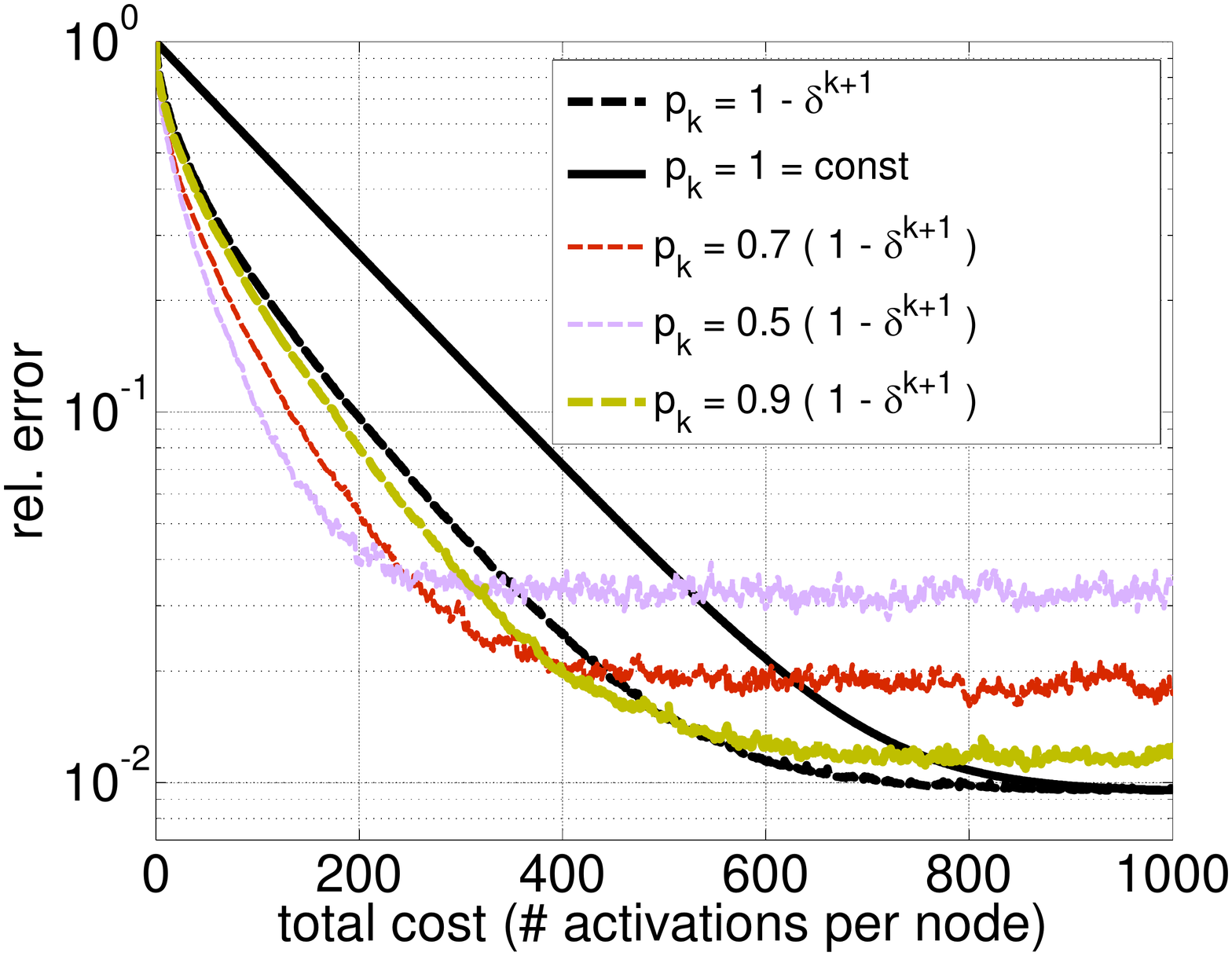}}
        %\caption{A gull}
        %\label{fig:gull}
        \caption{Relative error versus total cost (number of activations per node) for
      strongly convex quadratic costs, $n=40$-node network, and $\Lambda_i^k=0$.
      The Figures compare the following scenarios: 1) $p_k \equiv 1$ (standard DQN);
  2) $p_k=1-\sigma^{k+1}$;
  3) $p_k=p_{\mathrm{max}}\left(1-\sigma^{k+1}\right)$;
  and 4) $p_k = p_{\mathrm{max}}$, for all $k$.}
  \end{figure}

\begin{figure}
    \centering
    ~ %add desired spacing between images, e. g. ~, \quad, \qquad, \hfill etc.
    \subfigure[$\,$]{\includegraphics[height=2.6 in,width=3.2 in]{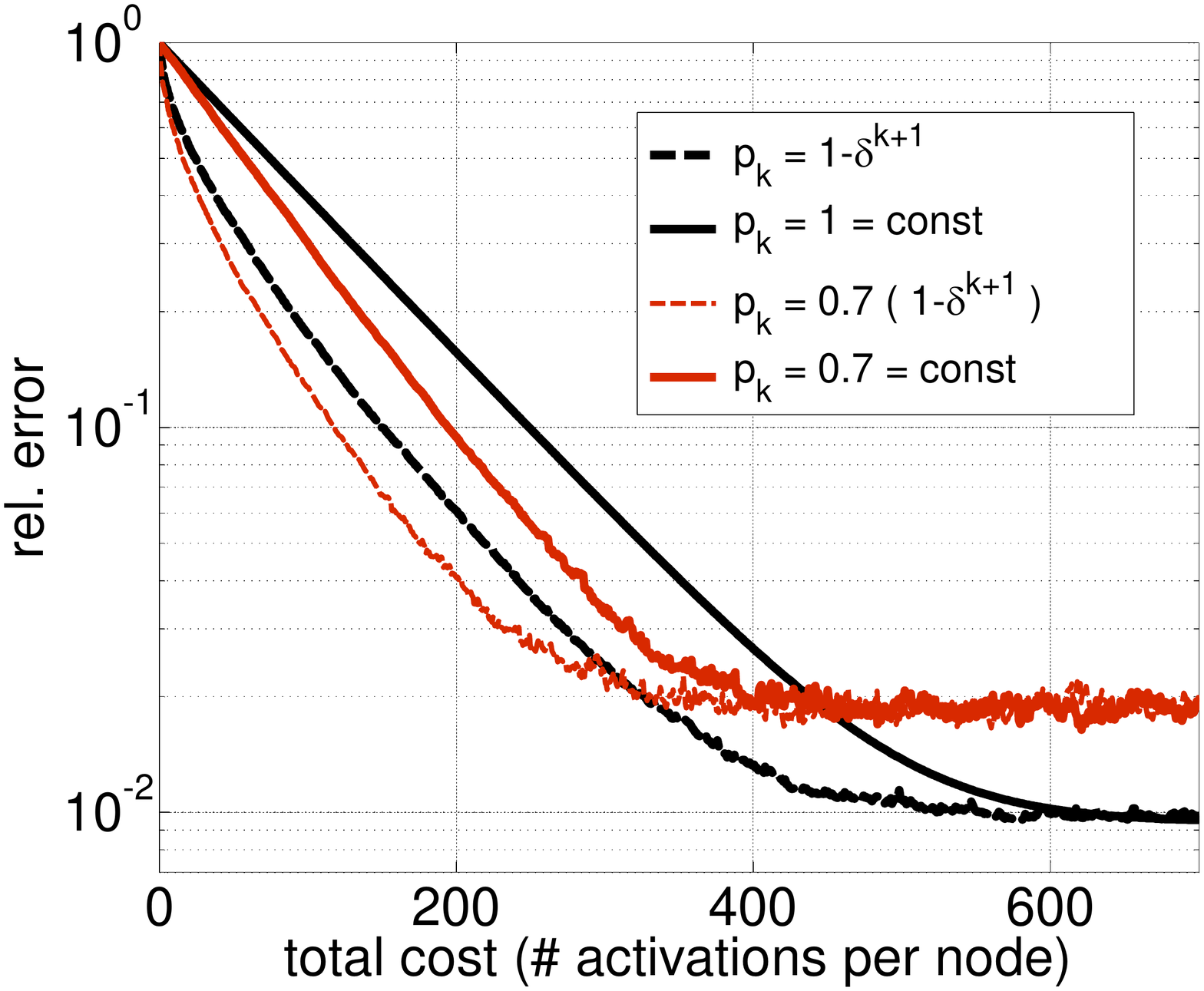}}
        %\caption{A gull}
        %\label{fig:gull}
    %\end{subfigure}
    \subfigure[$\,$]{\includegraphics[height=2.6 in,width=3.2 in]{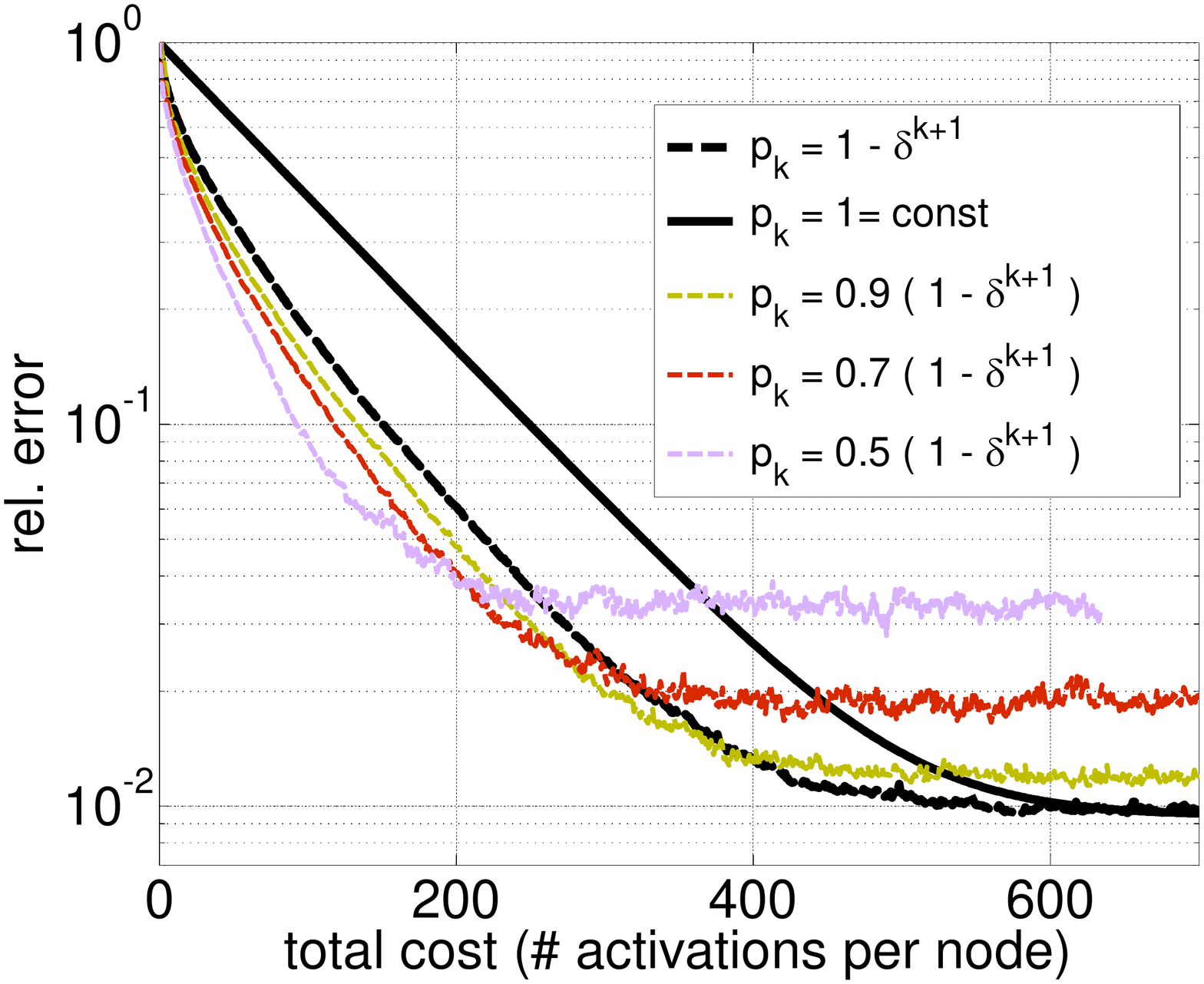}}
        %\caption{A gull}
        %\label{fig:gull}
   % \end{subfigure}
    \caption{Relative error versus total cost (number of activations per node) for
      strongly convex quadratic costs, $n=40$-node network, and $\Lambda_i^k=-I$.
      The Figures compare the following scenarios: 1) $p_k \equiv 1$ (standard DQN);
  2) $p_k=1-\sigma^{k+1}$;
  3) $p_k=p_{\mathrm{max}}\left(1-\sigma^{k+1}\right)$;
  and 4) $p_k = p_{\mathrm{max}}$, for all $k$.}
\end{figure}
%
%\begin{figure}[thpb]
%      \centering
%      \includegraphics[height=2. in,width=2.7 in]{SlikaNovo1.pdf}
%      \includegraphics[height=2. in,width=2.7 in]{SlikaNovo2.pdf}
%      \includegraphics[height=2. in,width=2.7 in]{SlikaNovo3.pdf}
%      \includegraphics[height=2. in,width=2.7 in]{SlikaNovo4.pdf}
%      %\includegraphics[height=2.9 in,width=3.4 in]{JournalFigFiner.pdf}
%      %\includegraphics[height=2.5in,width=3.62in]{slika_rate_p_3.pdf}
%      \caption{Relative error versus total cost (number of activations per node) for
%      quadratic costs and $n=40$-node network. First and second from top: $\Lambda_i^k=0$;
%      third and fourth from top: $\Lambda_i^k=-I$.
%      The Figures compare the following scenarios: 1) $p_k \equiv 1$ (standard DQN);
%  2) $p_k=1-\sigma^{k+1}$;
%  3) $p_k=p_{\mathrm{max}}\left(1-\sigma^{k+1}\right)$;
%  and 4) $p_k = p_{\mathrm{max}}$, for all $k$.}
%      \label{Figure_quadratic_small}
%\end{figure}

%\section*{Appendix}

\begin{figure}[thpb]
      \centering
      \subfigure[$\,$]{\includegraphics[height=3.2 in,width=2.7 in, angle = -90]{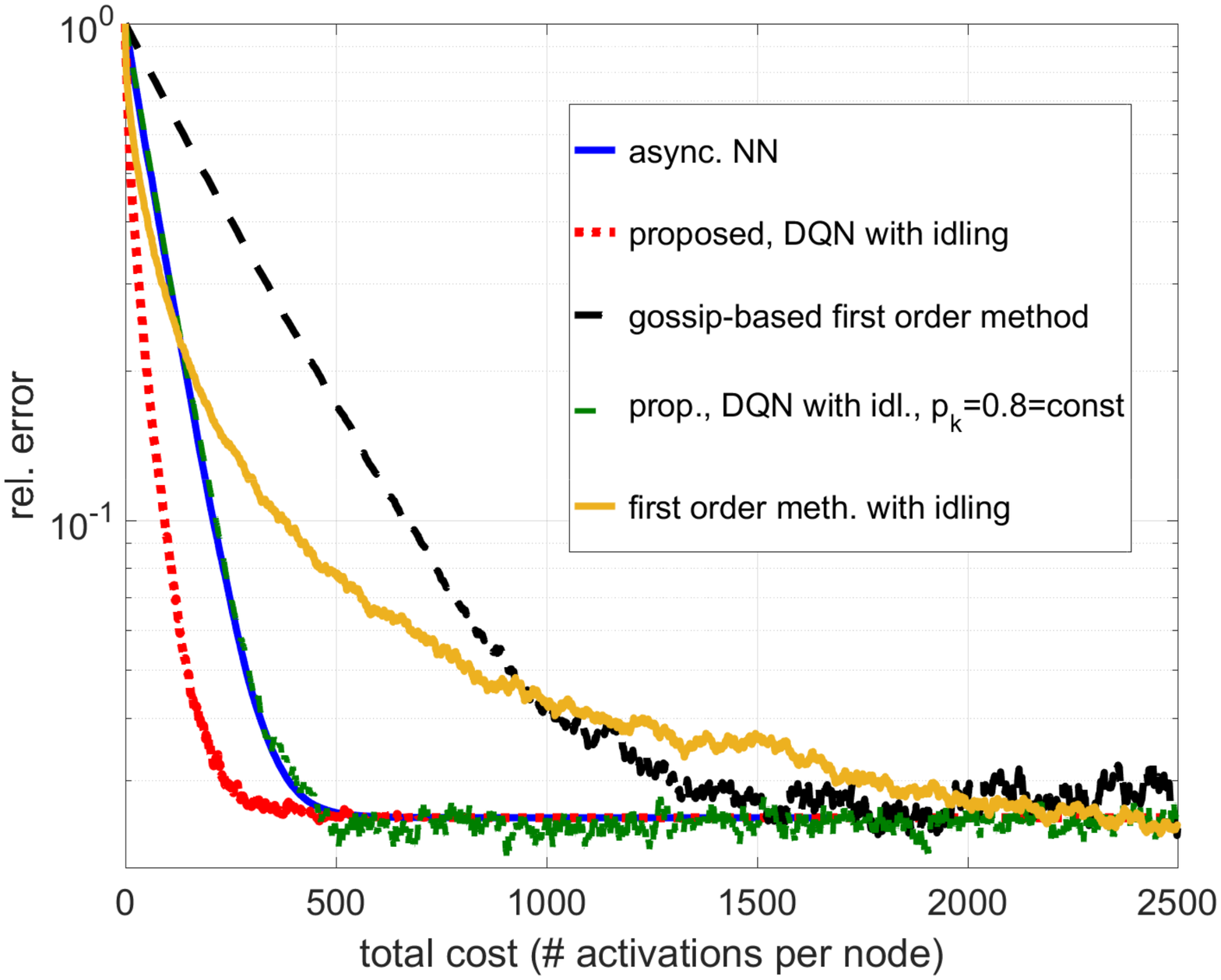}}
      \subfigure[$\,$]{\includegraphics[height=3.2 in,width=2.7 in, angle = -90]{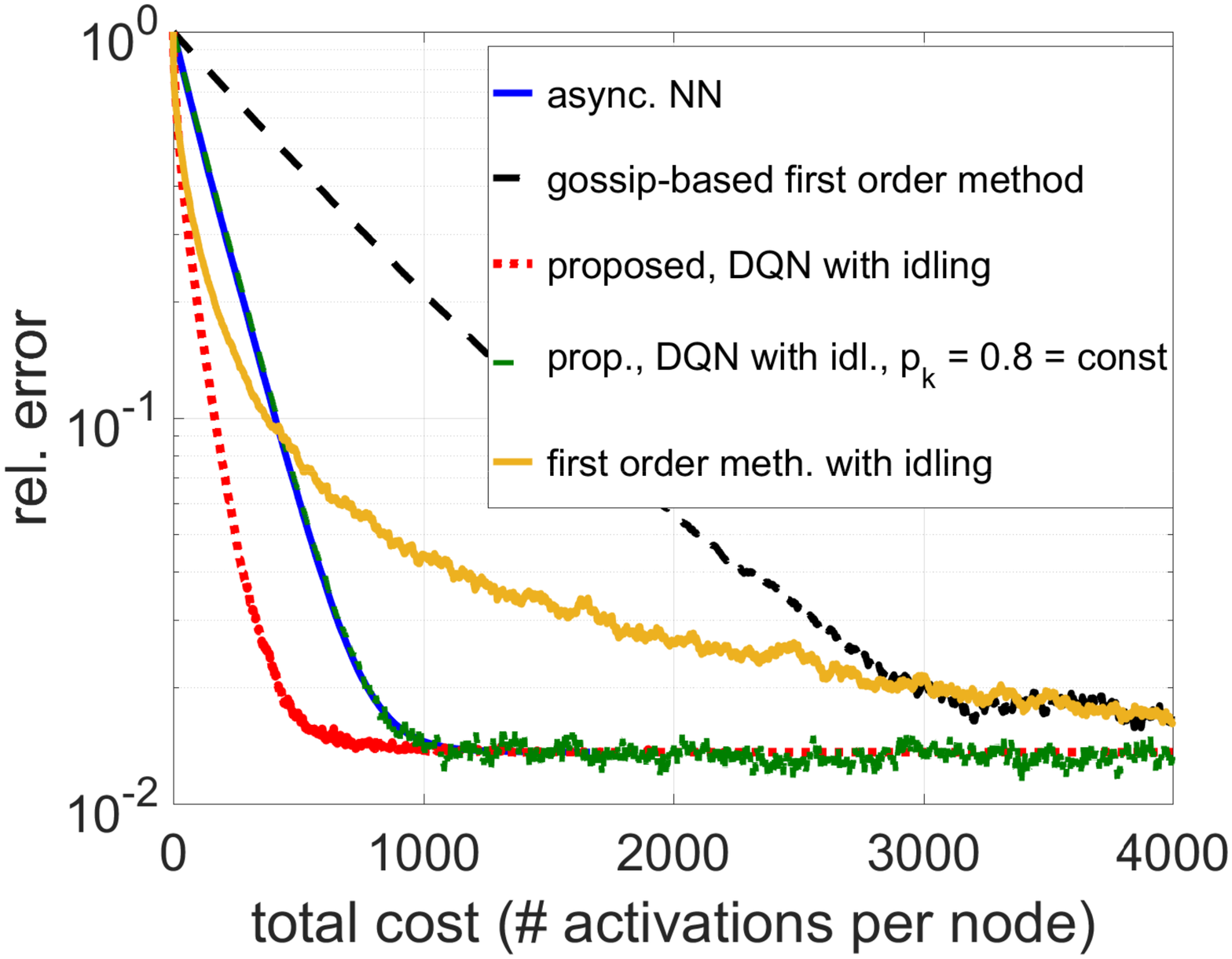}}
      \caption{Relative error versus total cost (number of activations per node) for
      strongly convex quadratic costs and $n=30$-node network and 
       $\alpha = 1/(100\,L)$ (top) and $\alpha = 1/(100\,L)$ (bottom). 
      The red, dotted line corresponds to the proposed DQN with idling
      and $p_k = 1-\sigma^{k+1}$;
      the yellow, solid line
      to the method in \cite{DGDSN} with
      $p_k = 1-((1-\alpha\,\mu)^2)^{k+1}$;
      blue, solid line to~\cite{ErminWeiAsyncNN};
      green, dashed line to the proposed DQN with idling
      and $p_k = 0.8 = \mathrm{const}$; and
      black, dashed line to the method in \cite{nedic-gossip}.}
      %\label{Figure_quadratic_small}
\end{figure}

We consider the problem with strongly convex local quadratic costs;
that is, for each $i=1,...,n$, we let
 $f_i:\,{\mathbb R}^p \rightarrow \mathbb R$,
 $f_i(x)=\frac{1}{2}(x-b_i)^T A_i (x-b_i)$,
  $p=10$,
 where $b_i \in {\mathbb R}^p$
  and $A_i \in {\mathbb R}^{p \times p}$
   is a symmetric positive definite matrix.
   The data pairs $A_i,b_i$ are
   generated at random, independently across nodes,
   as follows. Each $b_i$'s entry is
   generated mutually independently
   from the uniform distribution on~$[1,31]$.
   Each $B_i$ is generated as
   $B_i = Q_i \,D_i\,Q_i^T$; here, $Q_i$ is the
   matrix of orthonormal eigenvectors
   of~$\frac{1}{2}(\widehat{B}_i+\widehat{B}_i^T)$,
   and $\widehat{B}_i$ is a matrix
   with independent, identically distributed~(i.i.d.)
   standard Gaussian entries;
   and $D_i$ is a diagonal matrix
   with the diagonal entries drawn
   in an i.i.d. fashion from the uniform
   distribution on~$[1,31]$.

The network is a $n=100$-node
instance of the random geometric graph model
with the communication radius~$r =\sqrt{\frac{\mathrm{ln}(n)}{n}}$,
and it is connected.
The weight matrix~$W$ is set as follows:
 for~$\{i,j\} \in E$, $i\neq j$, $w_{ij}=\frac{1}{2(1+\max\{d_i,d_j\})}$,
 where $d_i$ is the node $i$'s degree;
 for~$\{i,j\} \notin E$, $i\neq j$, $w_{ij}=0$;
 and $w_{ii}=1-\sum_{j \neq i}w_{ij}$, for all~$i=1,...,n$.

We compare
the standard DQN method and the DQN method with
incorporated idling mechanism. Specifically,
we study how the relative error (averaged across nodes):
\[
\frac{1}{n}\sum_{i=1}^n \frac{\|x_i-\overline{x}^{*}\|}{\|\overline{x}^{*}\|},\,\,\overline{x}^{*} \neq 0,
\]
evolves with the elapsed
total number of activations per node.
Note that the number of activations
relates directly to both the communication
and computational costs of the algorithm.

The parameters for both algorithms
are set in the same way,
and the only difference is
in the activation schedule.
For the method with idling,
we set $p_k =1-\sigma^{k+1}$,
$k=0,1,...$
%***
 We set $\sigma = 1- c\,\alpha\,\mu$, with $c=40$.
%
%; how to set an optimal~$\delta$
% is an interesting future research direction.
%\footnote{For
%distributed first order
%method with idling,
%the parameter $\delta$
% can be optimally set
% based on the $f_i$'s
% strong convexity parameter and
% step size $\alpha$.
%  This can be done thanks to
%  the exact characterization of
%  the method's worst case
%  linear convergence factor~\cite{DGDSN}.
%  With distributed second order
%   methods considered here,
%   the analysis
%   is considerably more challenging,
%   and the established
%   linear convergence factor
%   for DQN with idling
%   is not claimed to be exact (tight).
%    For this reason it
%    is more challenging
%}
 (Clearly, for the method without
idling, $p_k \equiv 1$, for all~$k$.)
The remaining
algorithm parameters are as follows. We set
$\alpha=1/(100\,L)$,
where $L$ is the Lipschitz constant
of the gradients of the $f_i$'s
that we take as $\max_{i=1,...,n}\|A_i\|$.
 Further, we let $\theta=0$, and $\epsilon=1$ (full {step size}).
 %and $\rho=+\infty$ (no safeguarding).
 % [***I assumed here that
 %we will discuss the parameter choices and explain the parameters earlier in the paper.]
  We consider two choices for
 $\Lambda_i^k$ in step~6 of Algorithm~3:
 $\Lambda_i^k=0$, for all $i,k$;
 and $\Lambda_i^k=-I$, for all $i,k$.
 (We apply no safeguarding on the above choices of $\Lambda_i^k$, i.e.,
 we let $\rho=1$.)
 Note that the former choice
 corresponds to the algorithms
 with a single communication round per iteration~$k$,
 while the latter corresponds to the algorithms
 with two communication rounds per~$k$.

Figure~1 (a) plots the relative error versus
total cost (equal to total number of activations per node up to
the current iteration) for one sample path realization, for $\Lambda_i^k=0$.
We can see that incorporating the idling
mechanism significantly improves the
efficiency of the algorithm:
for the method to achieve the limiting
accuracy of approximately~$0.025$,
the method without idling takes about~$410$
 activations per node, while
 the method with idling takes
 about~$310$ activations. Hence,
  the idling mechanism
  reduces total cost by approximately~$24 \%$.
 Figure~1 (b)
  repeats the plots for
  $\Lambda_i^k=-I$, still showing
 clear gains of idling, though smaller than with $\Lambda_i^k=0$.

To account for randomness of the DQN method with idling that arises
due to the random nodes' activation schedule,
we include histograms of
the total cost needed to achieve a fixed
level of relative error. Specifically,
 in Figure~2 we plot histograms of the
 total cost (corresponding to $50$ generated
 sample paths -- $50$ different realizations of the $\xi_i^k$'s along iterations)
 needed to reach the relative error equal~$0.04$;
 Figure~2~(a) corresponds to $\Lambda_i^k=0$, while
 Figure~2~(b) corresponds to $\Lambda_i^k=-I$.
  The Figures also indicate with
  arrows the total cost needed
  by standard DQN to achieve the same accuracy.
   The results confirm the gains of idling. Also,
   the variability of total cost across
   different sample paths is small relative to
    the gain with respect to standard DQN.

Figures~3 and~4 investigate the scenarios
 when the activation probability~$p_k$
  may not asymptotically converge to one.
  The network is
  a (connected) random geometric graph
  instance with $n=40$ and $r=\sqrt{\frac{\mathrm{ln}(n)}{n}}$;
  step size~$\alpha=1/(200 L)$;
  the remaining system and algorithmic
  parameters are the same as with the previous simulation example.
  We consider
  the following
  choices for~$p_k$:
  1) $p_k \equiv 1$ (standard DQN);
  2) $p_k=1-\sigma^{k+1}$;
  3) $p_k=p_{\mathrm{max}}\left(1-\sigma^{k+1}\right)$;
  and 4) $p_k = p_{\mathrm{max}}$, for all $k$.
  With the third and
  fourth choices, the presence of $p_{\mathrm{max}}$
  models external effects on the $p_k$
   (out of control of the networked nodes), e.g.,
   due to link failures and unavailability
   of computing resources at certain iterations; it is varied within the set~$\{0.5,0.7,0.9\}$.
 Figure~3~(a)
  compares the methods for one sample path realization
  with the four choices of~$p_k$ above,
  with $p_{\mathrm{max}}=0.7$, for
  ${\Lambda_i^k}=0$.
  Figure~3~(b)
   compares for the same experiment the standard DQN ($p_k\equiv 1$),
     $p_k=1-\sigma^{k+1}$, and $p_k=p_{\mathrm{max}}\left(1-\sigma^{k+1}\right)$,
     with $p_{\mathrm{max}} \in \{0.5,0.7,0.9\}$.
      Several important observations stand out from the
      experiments. First,
      we can see that the limiting
      error increases when
      $p_k$ does not converge to one with respect
      to the case when it converges to one.
      However, this increase (deterioration)
       is moderate, and the algorithm
       still manages to converge to
        a good solution neighborhood
        despite the persisting idling.
        In particular,
        from Figure~3 (b),
        we can see that
        the limiting relative error
         increases from about~$10^{-2}$ (with $p_k \rightarrow 1$)
         to about~$3.5 \cdot 10^{-2}$
        with $p_{\mathrm{max}}=0.5$--a case
        with a strong persisting idling.
         This corroborates that DQN with idling
         is an effective method even when
         activation probability~$p_k$
          is not in full control of the algorithm designer.
          Second,
          the limiting error decreases
          when $p_{\mathrm{max}}$
           increases, as it is expected -- see Figure~3~(b).
           Finally,
           from Figure~3~(a),
            we can see that
            the method with $p_k=p_{\mathrm{max}}\left(1-\sigma^{k+1}\right)$
            performs significantly better than the
            method with $p_k=p_{\mathrm{max}}$, for all $k$.
            In particular, the methods
            have the same limiting error (approximately~$2 \cdot 10^{-2}$),
            while the former approaches this
            error much faster. This confirms
            that the proposed judicious design
            of increasing~$p_k$'s, as opposed
            to just keeping them constant,
            significantly improves the algorithm performance.
            Figure~4 repeats the same experiment for
             ${\Lambda_i^k}=-I$. We can see
             that the analogous conclusions can be drawn.

In the next experiment, we compare the
proposed DQN method with idling with
other existing methods that utilize randomized
activations of nodes. Specifically,
 we consider the very recent asynchronous (second order)
 network Newton method proposed in~\cite{ErminWeiAsyncNN}
 that is an asynchronous version of the method in~\cite{ribeiroNNpart1}.
 We refer to this method here as asynchronous Network Newton (NN).
 We also consider the
  first order method with idling proposed
  in~\cite{DGDSN}, and the
 first order gossip-based method in \cite{nedic-gossip}.
  The method in~\cite{ErminWeiAsyncNN} and the
DQN with idling and matrix $\mathbb{L} = -I$ both
utilize two $p$-dimensional communications per node
activation (they have equal communication cost per node activation).
The two methods also have a similar computational cost per activation.
 The method in~\cite{nedic-gossip}
 has a twice cheaper
 communication cost per activation,
 and it has in general a lower computational cost
 per activation (due to incorporating only
 the first order information into updates).

The comparison is carried out on a $n=30$-node
 (connected) random geometric graph instance with $91$ links,
 for the variable dimension $p=5$ and strongly convex
 quadratic $f_i$'s generated analogously to the previous experiments.
  With the proposed idling-DQN,
  we consider two choices of activation probabilities:
  1) $p_k = 1 - \sigma^{k+1}$, with
  $\sigma = 1- c\,\alpha\,\mu$, $c=40$, as in the previous experiment;
  and 2) $p_k=0.8 = \mathrm{const}$.
  With \cite{DGDSN},
  we let the activation probability $p_k = 1-((1-\alpha\,\mu)^2)^{k+1}$.
   The weight matrices of the proposed
  method and that in~\cite{ErminWeiAsyncNN,DGDSN}
   are set in the same way, as in prior experiments.
  With both idling-DQN and the method in \cite{ErminWeiAsyncNN},
  we set step size $\epsilon=1$.
   We consider two different choices of parameter~$\alpha$:
   $\alpha = 1/(100\,L)$, and $\alpha = 1/(200\,L)$.
     This is how step-sizes are set for
     the method in~\cite{ErminWeiAsyncNN} and for
     the idling-DQN with $p_k = 1 - \sigma^{k+1}$.
     Quantity $\alpha$ for the method in \cite{nedic-gossip} (the algorithm's step size)
     and for the idling-DQN with $p_k=0.8$
     are then adjusted (decreased) for a fair comparison, so that
     the four different methods achieve the same
     asymptotic relative error; we then look
     how many per-node activations each
     method takes to reach the saturating relative error.

 Figure~5 plots the relative error
 versus total number of activations
 for the four methods;
 Figure~(a) corresponds
 to $\alpha=1/(100\,L)$ and
 Figure~(b) is for $\alpha=1/(200\,L)$.
  We can see from Figure~5~(a) that
  the proposed idling-DQN with $p_k = 1-\sigma^{k+1}$
  outperforms the other methods:
  it takes about 400 per-node activations to
  reach the relative error $0.025$;
  the method in \cite{ErminWeiAsyncNN} and
  the idling-DQN with $p_k=0.8$ take about 600;
  and the method in \cite{nedic-gossip}
needs at least 1600 activations for the same accuracy.
 Note that, even when we half the number of activations
 for~\cite{nedic-gossip} to account for
 its twice cheaper communication cost,
 the proposed idling-DQN is still significantly faster --
 it compares versus~\cite{nedic-gossip} as 400 versus 800 ``normalized activations.''
 DQN with idling outperforms
 the method in~\cite{DGDSN} in terms of the number of
 activations, while it is to be noted
 that~\cite{DGDSN} has a lower computational cost per activation.
  Interestingly, the idling DQN with constant $p_k$ and
  the method in \cite{ErminWeiAsyncNN} practically
  match in performance.  Figure~5~(b) repeats the comparison
 for $\alpha = 1/(200\,L)$;
 we can see that similar conclusions can be drawn from this experiment.
  In summary, the idling-DQN
  reduces communication cost with
  respect to the method in~\cite{nedic-gossip},
  which is expected as it utilizes more (second order)
  computations at each activation.
  The two second order methods with
  randomized activations, the idling-DQN and the method in~\cite{ErminWeiAsyncNN},
  exhibit very similar performance when the idling-DQN uses constant~$p_k$ policy.
  With the increasing $p_k$ policy, the idling-DQN performs better than~\cite{ErminWeiAsyncNN}.
  This, together with previous experiments, demonstrates that a carefully designed workload orchestration with idling-DQN leads to performance improvements both with respect to a ``pure'' random activation policy and
  with respect to the all-nodes-work-all-time policy.

\section{Conclusion}
We incorporated an idling mechanism,
recently proposed in the context of
distributed first order methods~\cite{DGDSN},
into distributed second order methods.
Specifically, we study
the DQN algorithm~\cite{DQN}
 with idling.
% The idling mechanism
% involves
% an increasing number
% of nodes
% in the optimization process
% as the iteration counter~$k$
%  grows, so that, on average,
%  $n\,p_k$ nodes (out of $n$ nodes in total)
%  participate at iteration~$k$,
%  where $p_k$ is the per-node activation probability.
    We showed that, as long as $p_k$
    converges to one at least
      as fast as $1/k^{1+\zeta}$,
      $\zeta>0$ arbitrarily small,
    the DQN algorithm
    with idling converges
    in the mean square sense and almost surely
    to the same point
    as the standard DQN method
     that activates all nodes
     at all iterations. Furthermore,
      when $p_k$
       grows to one
       at a geometric rate,
       DQN with idling
       converges
       at a R-linear rate
       in the mean square sense.
        Therefore,
        DQN with idling
        achieves the same
        order of convergence (R-linear)
        as standard DQN, but with
        significantly cheaper iterations.
         Simulation examples
         corroborate communication and computational
         savings incurred by incorporating
         the idling mechanism
         and show the method's flexibility
         with respect to the choice
         of activation probabilities.

The proposed idling-DQN method,
and also other existing distributed second order methods
(involving local Hessian's computations)
with randomized nodes' activations, e.g., \cite{ErminWeiAsyncNN,RibeiroQN},
are not exact in the sense that they converge to
a solution neighborhood. An interesting future research
direction is to develop and analyze an idling-based
second order method with exact convergence.


\begin{thebibliography}{99}
%
%\bibitem{NocedalML}
%Bottou, L., Curtis, F., Nocedal, J., Optimization Methods for Large-Scale Machine Learning, 2016,
%available at: https://arxiv.org/abs/1606.04838

\bibitem{nedic_T-AC}
A. Nedi\'c, A. Ozdaglar, ``Distributed subgradient methods for multi-agent
  optimization,'' \emph{IEEE Transactions on Automatic Control},  vol.~54,
  no.~1, pp. 48--61, 2009.

 \bibitem{SayedEstimation}
F. Cattivelli, A. H. Sayed,  ``Diffusion {LMS} strategies for distributed
  estimation,'' \emph{IEEE Transactions on  Signal  Processing},  vol.~58, no.~3, pp.
  1035--1048, 2010.


\bibitem{WotaoYinExtra}
W. Shi, Q. Ling, G. Wu, W. Yin, ``{EXTRA}: an Exact First-Order Algorithm for Decentralized Consensus Optimization,''
\emph{SIAM Journal on Optimization}, vol. 2, no. 25, pp. 944-966, 2015.

\bibitem{ribeiroNNpart1} A. Mokhtari, Q. Ling, A. Ribeiro,
``Network {N}ewton Distributed Optimization Methods,''
\emph{IEEE Trans. Signal Processing}, vol. 65, no. 1, pp. 146-161, 2017.

%%%+++
\bibitem{BoydFusion}
L. Xiao, S. Boyd, S. Lall, ``A scheme for robust distributed sensor fusion
  based on average consensus,'' in \emph{IPSN '05, Information Processing in
  Sensor Networks}, 63--70, Los Angeles, California, 2005.

%%%+++
\bibitem{SoummyaSmartGrid}
G. Hug, S. Kar, C. Wu, ``Consensus + Innovations Approach for Distributed Multiagent Coordination in a Microgrid,'' \emph{IEEE Trans. Smart Grid}, vol. 6, no. 4, pp. 1893-1903, 2015.

%%%+++
\bibitem{BulloBook}
F. Bullo, J. Cortes, S. Martínez, ``Distributed Control of Robotic Networks:
{A} Mathematical Approach to Motion Coordination Algorithms,'' Princeton University Press, 2009.

%%%%+++
%\bibitem{Carlson}
%Carlson, David H., On real eigenvalues of complex matrices. Pacific J. Math. 15 (1965), no. 4, 1119--1129. http://projecteuclid.org/euclid.pjm/1102995269
%
%\bibitem{EigenvaluesNew}
%Zhang, F., Zhang, W., Eigenvalue Inequalities for Matrix Product, IEEE Trans. Aut. Contr. (2006), vol. 51, no. 9




%%%+++
\bibitem{DGDSN} D. Bajovi\'c, D. Jakoveti\'c, N. Kreji\'c, N. Krklec Jerinki\'c, ``Distributed Gradient Methods with Variable Number of Working Nodes,'' \emph{IEEE Trans. Signal Processing},
     vol. 64, no. 15, pp. 4080-4095, 2016.



\bibitem{schmidt} { M. P. Friedlander, M. Schmidt, } ``Hybrid
deterministic-stochastic methods for data fitting,'' \emph{ SIAM Journal on Scientific Computing},
vol. 34, pp. 1380-1405, 2012.

%%%+++
\bibitem{DQN} D. Bajovi\'c, D. Jakoveti\'c, N. Kreji\'c, N. Krklec Jerinki\'c, ``Newton-like method with  diagonal correction for distributed optimization,''
    \emph{SIAM J. Opt.}, vol. 27, no. 2, pp. 1171–1203. 2017.

%%%+++
\bibitem{NewtonRaphsonConsensus} D. Varagnolo, F. Zanella, A. Cenedese, G. Pillonetto, and L. Schenato,
``Newton-Raphson Consensus for Distributed Convex Optimization,''
\emph{IEEE Trans. Aut. Contr.}, vol. 61, no. 4, 2016.

%%%+++
\bibitem{RibeiroQN} M. Eisen, A. Mokhtari, A. Ribeiro, ``Decentralized quasi-{N}ewton methods,''
 \emph{IEEE Transactions on Signal Processing}, vol. 65, no. 10, pp. 2613--2628, May 2017.

    \bibitem{RibeiroQN2} M. Eisen, A. Mokhtari, A. Ribeiro,
    ``A Decentralized Quasi-Newton Method for Dual Formulations of
Consensus Optimization,'' \emph{IEEE 55th Conference on Decision and Control (CDC)},
 Las Vegas, NV, Dec. 2016.


        \bibitem{RibeiroQN3} M. Eisen, A. Mokhtari, A. Ribeiro, ``An asynchronous
        quasi-{N}ewton method for consensus optimization,''
        \emph{IEEE Global Conference on Signal and Information Processing (GlobalSIP)},
        Washington, DC, VA, USA, Dec. 2016.


        \bibitem{GlobalSip} D. Bajovi\'c, D. Jakoveti\'c, N. Kreji\'c, N. Krklec Jerinki\'c,
        ``Distributed first and second order
methods with variable number of working nodes,''
 \emph{IEEE Global
Conference on Signal and Information Processing},
Washington DC, VA, USA, Dec. 2016


%%%+++
\bibitem{DQM} A. Mokhtari, W. Shi, Q. Ling, A. Ribeiro,
``DQM: Decentralized Quadratically Approximated Alternating Direction Method of Multipliers,''
\emph{IEEE Trans. Signal Processing}, vol. 64, no. 19, pp. 5158-5173, 2016.

\bibitem{ErminWeiAsyncNN} F. Mansoori, E. Wei, ``Superlinearly Convergent Asynchronous Distributed Network Newton Method,'' 2017, available at https://arxiv.org/abs/1705.03952

%%%+++
\bibitem{cdc-submitted}
D. Jakoveti\'c,  J.~M.~F. Moura, J. Xavier, ``Distributed {N}esterov-like
  gradient algorithms'', \emph{CDC'12, 51$^\textrm{st}$ IEEE Conference on
  Decision and Control}, Maui, Hawaii, December 2012, pp. 5459--5464.

%%%+++
\bibitem{ESOM} A. Mokhtari, W. Shi, Q. Ling, A. Ribeiro,
``A Decentralized Second Order Method with Exact Linear Convergence Rate for Consensus Optimization,''
 \emph{IEEE Trans. Signal and Information Processing over Networks}, vol. 2, no. 4. pp. 507-522, 2016.

%%%++++
\bibitem{AccelDualAscent} M. Zargham, A. Ribeiro, A. Jadbabaie, ``Accelerated dual descent
for constrained convex network flow optimization,''
\emph{Decision and Control
(CDC), 2013 IEEE 52nd Annual Conference on}, Firenze, Italy, 2013. pp. 1037-1042.


%%%+++
  \bibitem{ASU_Math_Prog}
I. Lobel, A. Ozdaglar, D. Feijer, ``Distributed Multi-agent Optimization with State-Dependent Communication,''
\emph{Mathematical Programming,}  vol. 129, no. 2, pp. 255-284, 2014.

\bibitem{SayedAsync1} X. Zhao, A. H. Sayed, ``Asynchronous Adaptation and Learning Over Networks—--Part {I}: {M}odeling and Stability Analysis,''
    \emph{IEEE Transactions on Signal Processing}, vol. 63, no. 4, Feb. 2015.

\bibitem{SayedAsync2} X. Zhao, A. H. Sayed, ``Asynchronous Adaptation and Learning Over Networks—--Part {II}: {P}erformance Analysis,''
     \emph{IEEE Transactions on Signal Processing}, vol. 63, no. 4, Feb. 2015.

\bibitem{SayedWotaoYinAsync} T. Wu, K. Yuan, Q. Ling, W. Yin, A. H, Sayed, ``Decentralized Consensus Optimization with Asynchrony and Delays,''
    \emph{IEEE Transactions on Signal and Information Processing over Networks}, to appear, 2017, {DOI}: 10.1109/TSIPN.2017.2695121


%%%+++
\bibitem{WrightAsync} J. Liu, S. Wright, ``Asynchronous stochastic coordinate descent: {P}arallelism and convergence properties,'' \emph{SIAM J. Opt.}, vol. 25, no. 1, 2015.

%%%+++
\bibitem{HongADMM} T. H. Chang, L. Wei-Cheng, M. Hong, X. Wang, ``Distributed {ADMM} for large-scale optimization part~{II}: Linear convergence analysis and numerical performance,''
    \emph{IEEE Trans. Sig. Proc.}, vol. 64, no. 12, 2016.

%\bibitem{DuchiDelayed} Agarwal, A., Duchi, J., Distributed delayed stochastic optimization, in Advances in Neural Information Processing Systems, 2011.
%
% \bibitem{BoydADMM}
%Boyd, S., Parikh, N., Chu, E., Peleato, B., Eckstein, J., Distributed
%  optimization and statistical learning via the alternating direction method of
%  multipliers, Foundations and Trends in Machine Learning,
%Volume 3, Issue 1, (2011) pp. 1-122.

%
% \bibitem{noc1}  Byrd, R. H., Chin, G. M., Neveitt, W.,  Nocedal, J.,  On the Use of Stochastic Hessian Information in Optimization Methods for Machine Learning,
% \emph{ SIAM Journal on  Optimization,}  21 (3), (2011) pp. 977-995.
%
%\bibitem{noc2}  Byrd, R. H..  Chin, G. M.,  Nocedal, J.,  Wu, Y.,  Sample size selection in optimization methods for machine learning,
% {\em Mathematical Programming,} no. 134 vol. 1  (2012)   pp. 127-155.

% \bibitem{noc3}  Byrd, R. H., Hansen, S. L. ,  Nocedal, J.,  Singer, Y.,  A Stochastic Quasi-Newton Method for Large-Scale Optimization,
% {\em available at arXiv:1401.7020 [math.OC]}.

%%%+++


%\bibitem{AnnieChen}
%Chen I.-A., Ozdaglar, A.,  A fast distributed proximal gradient method, in
%  \emph{Allerton Conference on Communication, Control and Computing},
%  Monticello, {IL}, October 2012.
%
%
%\bibitem {ScutariBigData} Daneshmand, A., Facchinei, F., Kungurtsev, V., Scutari, G.,  Hybrid
%  random/deterministic parallel algorithms for nonconvex big data
%  optimization, \emph{submitted to IEEE Transactions on Signal Processing}, 2014.
%
%\bibitem{DES} Dembo, R.S. Eisenstat, S.C. Steinhaug, T., Inexact Newton method, \emph{SIAM Journall on  Numerical Analysis} 19,2, (1982), 400-409.

% \bibitem{FV} Feingold, D.G., Varga, R. S., Block Diagonally Dominant Matrices and Generalizations of the  Gershgorin  Circle
%Theorem, \emph{ Pacific Journal of Mathematics}, 12 (4), (1962) 1241-1250.
%
%
% \bibitem{schmidt} { Friedlander, M. P. , Schmidt, M., } Hybrid
%deterministic-stochastic methods \for data fitting, \emph{ SIAM Journal on Scientific Computing} 34
% (2012),  1380-1405.
%
%
%  \bibitem{arxivVersion}
%Jakoveti\'c, D., Xavier, J.,  Moura, J.~M.~F.,  Fast distributed gradient
%  methods, \emph{IEEE Transactions on  Automatic Control}, vol.~59, no.~5, (2014)  pp. 1131--1146.




%  \bibitem{SoummyaEst}
%Kar, S.,  Moura, J.~M.~F., Ramanan, K., Distributed parameter estimation in
%  sensor networks: Nonlinear observation models and imperfect communication,
%  \emph{IEEE Transactions on Information Theory},  vol.~58, no.~6, (2012) pp.
%  3575--3605.
%
%
% \bibitem{kk} { Kreji\'{c}, N.,
% Krklec, N., } Variable sample size methods for unconstrained
% optimization, \emph{ Journal of Computational and Applied Mathematics}  245 (2013),  213-231.

%%%+++
 \bibitem{kkj} {N.  Kreji\'{c}, N. Krklec Jerinki\'c, }
 ``Nonmonotone line search methods with
 variable sample size,''  \emph{Numerical Algorithms}, vol. 68, pp. 711--739, 2015.

% \bibitem{kl} Kreji\'c N., Lu\v{z}anin Z., Newton-like method with modification of the right-hand side vector, \emph{ Mathematics of Computation},  71, 237 (2002), 237-250.
%



%\bibitem{SayedConf}
%Lopes, C.,  {S}ayed,  A.~H., Adaptive estimation algorithms over distributed
%  networks, in \emph{21st IEICE Signal Processing Symposium}, Kyoto, Japan,
%  Nov. 2006.
%
%  \bibitem{ribeiro} Mokhtari, A., Ling, Q., Ribeiro, A.,
%An approximate Newton method for distributed optimization, 2014, available at http://www.seas.upenn.edu/~aryanm/wiki/NNICASSP.pdf

%%%+++


%%%%+++
%\bibitem{ribeiroNNpart2} Mokhtari, A., Ling, Q., Ribeiro, A., Network Newton--Part II: Convergence Rate and Implementation,
% 2015, available at: http://arxiv.org/abs/1504.06020

%\bibitem{ribeirobfgs1} Mokhtari, A., Ribeiro, A., Regularized Stochastic BFGS method, \emph{IEEE Transactions on Signal Processing}, Vol. 62, no. 23,  (2014) pp. 6089-6104.
%
%\bibitem{ribeirobfgs2} Mokhtari, A., Ribeiro, A., Global convergence of Online Limited Memory BFGS Method, J. Machine Learning Research, vol. (revised), April 2015.





%
%
%  \bibitem{JoaoMotaMPC}
%Mota, J., Xavier, J., Aguiar, P., P\"uschel, M., Distributed optimization
%  with local domains: {A}pplications in {MPC} and network flows, \emph{to
%  appear in IEEE Transactions on  Automatic Control}, 2015.

%%%+++


%%%+++
\bibitem{RamNedicVeeravalli}
S. Ram, A. Nedic, V. Veeravalli, ``Distributed stochastic subgradient
projection algorithms for convex optimization,''
\emph{J. Optim. Theory Appl.},
vol. 147, no. 3, pp. 516–-545, 2011.

%\bibitem{nedic_novo}
%Ram, S., Nedi\'c, A., Veeravalli, V., Distributed stochastic subgradient
%  projection algorithms for convex optimization, \emph{Journal on  Optimization Theory and
%  Applications},  vol. 147, no.~3, (2011) pp. 516--545.

%%%+++
\bibitem{nedic-gossip}
S. S. Ram, A. Nedi\'c, V. Veeravalli, ``Asynchronous gossip algorithms for
  stochastic optimization,'' \emph{CDC '09, 48th IEEE International
  Conference on Decision and Control}, Shanghai, China, December 2009, pp. 3581
  -- 3586.


%
%\bibitem{RibeiroADMM1}
%Schizas, I.~D. , Ribeiro, A.,  Giannakis,  G.~B., Consensus in ad hoc {WSN}s
%  with noisy links -- {P}art~{I}: {D}istributed estimation of deterministic
%  signals, \emph{IEEE Transactions on  Signal Processing},  vol.~56, no.~1, (2009) pp. 350--364.
%
%
%\bibitem{IonMatei} Matei, I., Baras, J.,
%Performance evaluation of the consensus-based distributed subgradient method under random communication topologies, IEEE Journal of Selected Topics in Signal Processing 5 (4), 754-771, 2011.
%
%\bibitem{JohanssonMaximum}
%Shi, G.,   Johansson, K. H., Finite-time and asymptotic convergence
%of distributed averaging and maximizing algorithms, 2012, available at:
%http://arxiv.org/pdf/1205.1733.pdf
%
%\bibitem{DadmmLinear} W. Shi, Q. Ling, K. Yuan, G. Wu, and W. Yin,
%On the Linear Convergence of the ADMM in Decentralized Consensus Optimization,
%IEEE Transactions on Signal Processing vol. 62, no. 7, 2013.

%%%+++





%%%+++
\bibitem{EWnew}  E. Wei, A. Ozdaglar, A. Jadbabaie, ``A distributed Newton method
for network utility maximization–--{I}: {A}lgorithm,'' \emph{ IEEE
Transactions on Automatic Control,} vol. 58, no. 9, pp. 2162- 2175, 2013.






%\bibitem{WotaoYinDisGrad}
%Yuan, L., Ling, Q., Yin, W., On the convergence of decentralized gradient
%  descent, 2013, available at: http://arxiv.org/abs/1310.7063.




%\bibitem{NNDDtsp} Bajovic, D. Jakovetic, D., Krejic, N., Krklec Jerinkic, N., Distributed Gradient Methods with Variable Number of Working Nodes,
%  to appear in IEEE Transactions on Signal Processing, 2016.

 \end{thebibliography}
\end{document}